\documentstyle[psfig,epsfig]{mn}

\newif\ifAMStwofonts

\newcommand{\flex}{\mbox{$\mathcal{F}$}}
\newcommand{\sflex}{\mbox{$\mathcal{G}$}}

\newcommand{\dif}{\mbox{$\mathrm{d}$}}

\newcommand{\thetab}{\mbox{\boldmath $\theta$}}
\newcommand{\nablab}{\mbox{\boldmath $\nabla$}}

\newcommand{\mi}{\mathrm{i}}

\ifoldfss
  \ifCUPmtlplainloaded \else
    \NewTextAlphabet{textbfit} {cmbxti10} {}
    \NewTextAlphabet{textbfss} {cmssbx10} {}
    \NewMathAlphabet{mathbfit} {cmbxti10} {} 
    \NewMathAlphabet{mathbfss} {cmssbx10} {} 
  \fi
  \ifAMStwofonts
    \ifCUPmtlplainloaded \else
      \NewSymbolFont{upmath} {eurm10}
      \NewSymbolFont{AMSa} {msam10}
      \NewMathSymbol{\upi}     {0}{upmath}{19}
      \NewMathSymbol{\umu}     {0}{upmath}{16}-----------------------------------------------------------------------

      \NewMathSymbol{\upartial}{0}{upmath}{40}
      \NewMathSymbol{\leqslant}{3}{AMSa}{36}
      \NewMathSymbol{\geqslant}{3}{AMSa}{3E}

    \fi
  \fi
\fi 

\ifnfssone
  \newmathalphabet{\mathit}
  \addtoversion{normal}{\mathit}{cmr}{m}{it}
  \addtoversion{bold}{\mathit}{cmr}{bx}{it}
  \newmathalphabet{\mathbfit} 
  \addtoversion{normal}{\mathbfit}{cmr}{bx}{it}
  \addtoversion{bold}{\mathbfit}{cmr}{bx}{it}
  \newmathalphabet{\mathbfss} 
  \addtoversion{normal}{\mathbfss}{cmss}{bx}{n}
  \addtoversion{bold}{\mathbfss}{cmss}{bx}{n}
  \ifAMStwofonts
    \ifCUPmtlplainloaded \else
      \UseAMStwoboldmath
      \makeatletter
      \new@mathgroup\upmath@group
      \define@mathgroup\mv@normal\upmath@group{eur}{m}{n}
      \define@mathgroup\mv@bold\upmath@group{eur}{b}{n}
      \edef\UPM{\hexnumber\upmath@group}
      \new@mathgroup\amsa@group
      \define@mathgroup\mv@normal\amsa@group{msa}{m}{n}
      \define@mathgroup\mv@bold\amsa@group{msa}{m}{n}
      \edef\AMSa{\hexnumber\amsa@group}
      \makeatother
      \mathchardef\upi="0\UPM19
      \mathchardef\umu="0\UPM16
      \mathchardef\upartial="0\UPM40
      \mathchardef\leqslant="3\AMSa36
      \mathchardef\geqslant="3\AMSa3E
    \fi
  \fi
\fi 

\ifnfsstwo
  \DeclareMathAlphabet{\mathbfit}{OT1}{cmr}{bx}{it}
  \SetMathAlphabet\mathbfit{bold}{OT1}{cmr}{bx}{it}
  \DeclareMathAlphabet{\mathbfss}{OT1}{cmss}{bx}{n}
  \SetMathAlphabet\mathbfss{bold}{OT1}{cmss}{bx}{n}
  \ifAMStwofonts
    \ifCUPmtlplainloaded \else
      \DeclareSymbolFont{UPM}{U}{eur}{m}{n}
      \SetSymbolFont{UPM}{bold}{U}{eur}{b}{n}
      \DeclareSymbolFont{AMSa}{U}{msa}{m}{n}
      \DeclareMathSymbol{\upi}{0}{UPM}{"19}
      \DeclareMathSymbol{\umu}{0}{UPM}{"16}
      \DeclareMathSymbol{\upartial}{0}{UPM}{"40}
      \DeclareMathSymbol{\leqslant}{3}{AMSa}{"36}
      \DeclareMathSymbol{\geqslant}{3}{AMSa}{"3E}
    \fi
  \fi
\fi 

\ifCUPmtlplainloaded \else
  \ifAMStwofonts \else 
    \def\upi{\pi}
    \def\umu{\mu}
    \def\upartial{\partial}
  \fi
\fi
\newcommand{\be}{\begin{equation}}
\newcommand{\ee}{\end{equation}}
\newcommand{\ba}{\begin{eqnarray}}
\newcommand{\ea}{\end{eqnarray}}
\newcommand{\nn}{\nonumber \\}

\newcommand{\de}{\partial}

\newcommand{\atanh}{\mbox{$\mathrm{arctanh}$}}

\title[Weak Gravitational Flexion] {Weak Gravitational Flexion} \author[D. J. Bacon et al]
{D. J. Bacon$^{1*}$, D. M. Goldberg$^{2}$, B. T. P. Rowe$^1$, A. N. Taylor$^1$\\ $^1$Institute for Astronomy,
Royal Observatory Edinburgh, Blackford Hill, Edinburgh, EH9 3HJ,
U. K.\\ $^2$ Department of Physics, Drexel University, Philadelphia,
PA 19104, USA \\ $^*$email: djb@roe.ac.uk} \date{}

\pagerange{\pageref{firstpage}--\pageref{lastpage}}
\pubyear{2004}

\begin{document}

\maketitle

\label{firstpage}

\begin{abstract}
Flexion is the significant third-order weak gravitational lensing
effect responsible for the weakly skewed and arc-like appearance of
lensed galaxies. Here we demonstrate how flexion measurements can be
used to measure galaxy halo density profiles and large-scale structure
on non-linear scales, via galaxy-galaxy lensing, dark matter mapping
and cosmic flexion correlation functions. We describe the origin of
gravitational flexion, and discuss its four components, two of which
are first described here. We also introduce an efficient complex
formalism for all orders of lensing distortion. We proceed to examine
the flexion predictions for galaxy-galaxy lensing, examining
isothermal sphere and Navarro, Frenk \& White (NFW) profiles and both
circularly symmetric and elliptical cases. We show that in combination
with shear we can precisely measure galaxy masses and NFW halo
concentrations. We also show how flexion measurements can be used to
reconstruct mass maps in 2-D projection on the sky, and in 3-D in
combination with redshift data. Finally, we examine the predictions
for cosmic flexion, including convergence-flexion cross-correlations,
and find that the signal is an effective probe of structure on
non-linear scales.
\end{abstract}


\section{Introduction}

Weak gravitational lensing is a rapidly developing subject, with great
progress being made in many related observational areas. The mass and
density profiles of galaxies have been carefully explored using
galaxy-galaxy shear studies (e.g. Hoekstra et al 2004), while
large-scale structure can be traced using cosmic shear (see e.g. van
Waerbeke \& Mellier 2003, Refregier 2003 for reviews). This has led to
significant constraints on cosmological parameters, such as the
fluctuation of the matter distribution, the density of matter, and the
growth rate of matter fluctuations in the Universe.

Gravitational lensing has received so much interest partially because
it allows us to measure the mass of structures with very few physical
assumptions. The distortion of background galaxies depends only on the
geometry of the lens system, the mass, and the use of the weak-field
limit of General Relativity. As such, lensing presents us with a
method for measuring mass which is free of dynamical uncertainties
associated with questions as to whether the system is relaxed. It is a
direct measure of the mass present, whether in visible or dark form.

Weak gravitational lensing is typically studied by examining the
ellipticities of source galaxies, seeking a coherent alignment of
these ellipticities (or other combinations of weighted second-order
moments of galaxy light) induced by mass along the line of sight
(e.g. Kaiser, Squires \& Broadhurst, 1995, Kaiser 2000, Bernstein \&
Jarvis 2002, Refregier \& Bacon 2003, Hirata \& Seljak 2003). However,
Goldberg \& Natarajan (2002) have shown that significant further
information is available from the skewedness and arciness of the light
distribution for source galaxies; we have further developed this
approach in Goldberg \& Bacon (2005) where we have labelled this third
order effect as the ``flexion'' of these images. A related approach
using `sextupole lensing' has recently been explored by Irwin \&
Shmakova (2005).

In our previous paper (Goldberg \& Bacon 2005), we described the
theory of flexion, and demonstrated how this effect can be measured
using the Shapelet formalism (Bernstein \& Jarvis 2002, Refregier
2003, Refregier \& Bacon 2003). We also demonstrated that the flexion
signal is present in Deep Lens Survey data (Wittman et al 2002).

In this paper, we explore and describe what flexion is able to teach
us in the context of several cosmological applications: how flexion
can contribute to our understanding of galaxy mass and density
profiles; its usefulness in creating maps of the dark matter
distribution; and its value for measuring large-scale structure in the
non-linear regime.

In Section 2, we will give a brief introduction to the flexion
formalism, and will revise the process by which flexion is measured
using shapelets. Section 3 introduces two forms of flexion, one of
which was was not discussed in our previous work; we find that both
forms of flexion are a high-pass filter for projected density
fluctuations, with one form of flexion measuring local information
about density, and the other measuring non-local information. Here we
also discuss whether flexion as presented so far is an efficient
description of arced galaxy shapes.

Section 4 examines flexion predictions for galaxy-galaxy lensing,
concentrating on averaged circular profiles; we discuss how flexion
can be used to provide more information about galaxy profiles, and how
combination of the flexion with shear can break mass-sheet
degeneracies. Section 5 extends this analysis to elliptical density
profiles. 

In Section 6 we show how flexion can be used for mass reconstruction,
and note the utility of flexion for measuring substructure in
clusters. Section 7 discusses the use of flexion for measurements of
large-scale structure; we find that the cosmic flexion signal is 
measurable exclusively on non-linear scales, which are nevertheless of
great interest. We conclude in Section 8.

\section{Flexion Formalism}

We begin by briefly reviewing the flexion formalism as developed by
Goldberg \& Bacon (2005), examining how flexion is defined and how it
can be measured using shapelets.

\subsection{Flexion}

It is useful to start by noting the importance in lensing of the
dimensionless surface density of matter, the convergence
$\kappa$. This is defined for a set of source objects at angular
diameter distance $D_s$, which have been lensed by a mass at angular
diameter distance $D_l$. Then
\begin{equation}
\kappa({{\bmath \theta}})\equiv \frac{D_{ls} D_l}{D_s} \frac{4\pi G
\Sigma({\bmath \theta})}{c^2}\ ,
\label{eq:kappa_def}
\end{equation}
with ${\bmath \theta}$ the image coordinates for the observer, and
$\Sigma$ the projected surface density of the lens.

The relationship between unlensed coordinates and lensed, observed
coordinates is given by
\begin{eqnarray}
A_{ij}({{\bf \theta}})&\equiv&
\frac{\partial{{\bf \theta}_i'}}{\partial {{\bf \theta}_j}}=
\left( 
\delta_{ij}-\partial_i \partial_j \psi({{\bf \theta}})\right),\\\ \nonumber
{\bf A}&=&
\left(
\begin{array}{cc}
1-\kappa-\gamma_1 & -\gamma_2 \\
-\gamma_2 & 1-\kappa+\gamma_1
\end{array}
\right)\ 
\label{eq:Adef}
\end{eqnarray}
where $\partial_i\equiv \partial/\partial \theta_i$, and ${\bf
\theta}'$ are the unlensed coordinates; the origins of the measured,
lensed coordinates and the unlensed source coordinates are taken to be
the centres of light for the lensed and unlensed images
respectively. Here $\psi$ is the lensing potential, i.e. a projected
gravitational potential along the line of sight.

If convergence and shear are effectively constant within a source galaxy
image, the galaxy's transformation can simply be described as:
\begin{equation}
\theta'_i=A_{ij}\theta_j\ .
\end{equation}
Flexion arises from the fact that the shear and convergence are
actually not constant within the image, and so we have to expand to
second order:

\begin{equation}
\theta'_i\simeq A_{ij}\theta_j +\frac{1}{2}D_{ijk} \theta_j \theta_k,
\end{equation}
with 
\begin{equation}
D_{ijk}=\partial_k A_{ij}.
\end{equation}
Using results from Kaiser (1995), we find that
\begin{eqnarray}
D_{ij1}&=&\left(
\begin{array}{cc}
-2\gamma_{1,1}-\gamma_{2,2} & \ \ -\gamma_{2,1} \\
-\gamma_{2,1} & \ \ -\gamma_{2,2} 
\end{array}
\right), \\ \nonumber
D_{ij2}&=& \left(
\begin{array}{cc}
-\gamma_{2,1} & \ \ -\gamma_{2,2}\\
-\gamma_{2,2} & \ \ 2\gamma_{1,2}-\gamma_{2,1}
\end{array}
\right).
\label{eq:ddef}
\end{eqnarray}
By expanding the surface brightness as a Taylor series and using the
relations above, we find that we can approximate the lensed surface
brightness of a galaxy in the weak lensing regime as
\begin{equation}
f({\bf \theta})\simeq\left\{1+
	\left[ (A-I)_{ij}\theta_j+\frac{1}{2}D_{ijk}\theta_j\theta_k\right]
	\partial_i \right\}	f'({\bf \theta}) \ .
\label{eq:2ndorder}
\end{equation}
This shows that the flexion lensing effects are in terms of
derivatives of the shear field. We define the flexion in terms of
these shear derivatives, using the combination which is shown by Kaiser (1995)
to give the gradient of the convergence:

\begin{eqnarray}
{\bf {\cal F}}&\equiv& (\gamma_{1,1}+\gamma_{2,2}){\bf
  i}+
(\gamma_{2,1}-\gamma_{1,2}){\bf j}
\label{eq:f1def}\nn
&=&\nabla \kappa\nn
&=&|{\cal F}| e^{i\phi}.
\label{eq:f1def}
\end{eqnarray}
Since the flexion is in terms of derivatives of the shear field, we
therefore require a means of measuring these derivatives,
$\gamma_{i,j}$.

\subsection{Shapelet Measurement}

We have found (Goldberg \& Bacon 2005) that we can measure derivatives
of the shear, and hence obtain measurements of the flexion, using the
shapelet formalism of Refregier (2003) and Bernstein \& Jarvis (2002),
as applied to lensing by Refregier \& Bacon (2003).

We decompose galaxy images into shapelet coefficients, corresponding to 
prefactors for reduced Hermite polynomials:

\begin{equation}
f({\bf \theta})=\sum_{n,m} f_{nm} B_{nm}({\bf \theta}) \ 
\end{equation}
where
\begin{equation}
B_{nm}({\bf \theta};\beta)=\beta^{-1}\phi_n(\beta^{-1}\theta_1)\phi_m(\beta^{-1}\theta_2)\ .
\end{equation}
Here $\beta$ is a scale factor chosen for the galaxy, and $\phi_n$ are
reduced Hermite polynomials.

Since these functions are eigenfunctions for the quantum harmonic
oscillator, we can define ladder operators as in quantum mechanics:
\begin{eqnarray}
\hat{a}_1 \left| \phi_{n\ m}\right>&=&\sqrt{n} \left| \phi_{n-1\  m}
\right> \nn
\hat{a}^\dagger_1 \left| \phi_{n\ m}\right>&=&\sqrt{n+1} \left|
\phi_{n+1\ m} \right>
\end{eqnarray}
and describe lensing distortions in terms of these
operators. Explicitly, we find that the lensed image intensity is
given by:

\begin{equation}
f({\bf \theta})\simeq (1+\kappa \hat{K}+\gamma_i
\hat{S}^{(1)}_i+\gamma_{i,j}\hat{S}^{(2)}_{ij}) f({\bf \theta}')
\label{eq:lens}
\end{equation}
where each lensing operator, including the $S_{ij}$ second order
lensing effect, is given in terms of $a$ and $a^\dagger$. The explicit
forms are somewhat complex, and are given in full in Goldberg \& Bacon
(2005). We also show in that paper that the second-order lensing induces
a shift in the centroid of an object, and give explicit forms for this
shift.

We measure $\gamma_{i,j}$ by $\chi^2$ fitting to a version of
equation~(\ref{eq:lens}), simplified by the lack of cross-talk between
odd and even shapelet coefficients (see Goldberg \& Bacon 2005 for
details).  Then from the estimated shear derivatives, we can calculate the
flexion according to equation~(\ref{eq:f1def}).

In addition, Goldberg \& Bacon (2005) have measured the shapelet
coefficients and derive flexion and shear for 4833 pairs of galaxies
in the Deep Lens Survey.  We find that using flexion alone, the
averaged lens galaxy may be fit by an isothermal sphere with a
characteristic velocity width of 220 km/s.  Having established in that
paper that the flexion signal is indeed measurable, we devote this
work to developing new flexion analysis techniques.

\section{Complex Representation and Second Flexion}

In this section we develop a compact and straightforward complex
formalism for flexion, which is of much wider applicability to all weak
gravitational lensing. In addition we show that weakly lensed arcs can be
uniquely decomposed into the spin-1 first flexion of Section 2, and a
new component which has not previously been considered, the second
flexion which we show has spin-3 properties. We begin by re-deriving
the shear in complex notation.

We define a complex gradient operator: 
\be 
\de = \de_1 + i \de_2, 
\ee
which we can think of as a derivative with an amplitude and a
direction down the slope of a surface at any point. It transforms
under rotations as a vector, $\de'=\de e^{i \phi}$, where $\phi$ is
the angle of rotation. This operator can be compared with the
covariant derivative formalism of Castro et al (2005) for weak lensing
on the curved sky. Applying the operator to the lensing scalar
potential, $\psi$, we can generate the spin-1 (i.e. vector) lensing
displacement field,

\be 
\alpha = \alpha_1 + i \alpha_2 = \de \psi.  
\ee
This correspondence allows us to interpret the complex gradient,
$\de$, as a spin-raising operator, raising the function it acts on by
one spin value. Similarly the spin of a quantity can be lowered by
applying the complex conjugate gradient, $\de^*$.  Applying one after
the other yields the spin-zero 2-D Laplacian, 
\be 
\de \de^* = \de^*\de, 
\ee 
where we have noted that $\de$ and $\de^*$ commute.  Applying the
complex conjugate derivative to the displacement field we find the spin is
lowered to the spin-0 convergence field 
\be 
\kappa = \frac{1}{2} \de^*\alpha = \frac{1}{2} \de^* \de \psi.  
\ee
Applying the spin-raising operation to the displacement field raises
us to a spin-2 field, the complex shear: 
\be \gamma = \gamma_1 + i\gamma_2 = \frac{1}{2}\de \de \psi.
\label{eqn:gamma} 
\ee 
From these expressions it is easy to recover the general, complex
Kaiser-Squires (1993) relation between the shear and convergence
fields, 
\be 
\kappa + i B = \de^{-2} \de^* \de^* \gamma,
\label{cmplxks} 
\ee where $\de^{-2}$ is the 2-dimensional inverse Laplacian, and the
non-lensing, curl/odd-parity $B$-field is automatically included as
the complex part of the recovered field. We can also see from this
relation that a $B$-field can be generated from a convergence field by
a $\pi/4$ rotation of the shear field, equivalent to multiplying the
complex shear by $i$. In equation (\ref{cmplxks}) we have omitted an
arbitrary constant, due to the sheet-mass degeneracy.

\begin{figure}
\psfig{figure=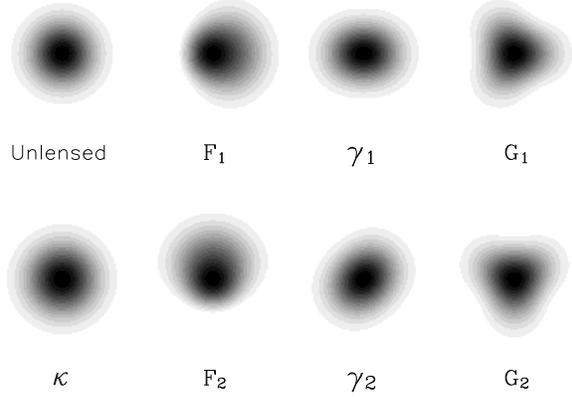,angle=0,width=8cm}
\caption{Weak lensing distortions with increasing spin values. Here an
unlensed Gaussian galaxy with radius 1 arcsec has been distorted with
10\% convergence/shear, and 0.28 arcsec$^{-1}$ flexion. Convergence
is a spin-0 quantity; first flexion is spin-1; shear is spin-2; and
second flexion is spin-3.}
\label{fg:eigenspace}
\end{figure}

The complex formalism provides a neat way to generalize the
analysis of distortions to higher orders. Taking the third
derivative of the lensing potential we have the unique combinations
 \ba
    {\cal F}    &=& |{\cal F}|  e^{i \phi} =
    \frac{1}{2} \de \de^* \de \psi = \de \kappa = \de^* \gamma,\nn
    {\cal G}    &=& |{\cal G}| e^{i3 \phi} = \frac{1}{2} \de \de \de \psi = \de \gamma,
\label{eqn:dfg}
 \ea
where the first flexion, ${\cal F}$, is a spin-1 field and the new
second flexion,  ${\cal G}$, is seen to be a spin-3 field. Here $\phi$
represents the position angle determining the direction of the vector
or spin-3 component.
Expanding the flexions in terms of the gradients of the shear field we
find
 \ba
{\cal F}    &=& (\de_1 \gamma_1 + \de_2 \gamma_2) + i (\de_1
                \gamma_2-\de_2 \gamma_1)\nn
{\cal G}    &=& (\de_1 \gamma_1 - \de_2 \gamma_2) + i (\de_1
                \gamma_2+\de_2 \gamma_1),
\label{eqn:fg}
 \ea
where the definition of the first flexion agrees with our previous
results in Section 2. These two independent fields
specify the weak ``arciness'' of the lensed image.

The complex representation allows us to find a consistency relation
between the two flexion fields, 
  \be
    \de^* \de {\cal G} = \de \de {\cal F},
  \ee
which can be used as a check on measurements of ${\cal F}$ and
${\cal G}$.

We are also able to obtain a direct
description of the third order lensing tensor $D_{ijk}$. Defining
$\flex = \flex_{1} + \mi \flex_2$ and $\sflex = \sflex_1 + \mi
\sflex_2$ we can then re-express $D_{ijk}$ as the sum of two terms
$D_{ijk} = \flex_{ijk} + \sflex_{ijk}$, where the first (spin-1) term is
\begin{eqnarray}
\flex_{ij1}&=&-\frac{1}{2}\left(
\begin{array}{cc}
3 \flex_1 & \ \ \flex_2 \\
\flex_2 & \ \ \flex_1 
\end{array}
\right) \\ \nonumber
\flex_{ij2}&=&-\frac{1}{2} \left(
\begin{array}{cc}
\flex_2  & \ \ \flex_1 \\
\flex_1  & \ \ 3\flex_2
\end{array}
\right)
\label{eq:fdef}
\end{eqnarray}
and the second (spin-3) term is
\begin{eqnarray}
\sflex_{ij1}&=&-\frac{1}{2} \left(
\begin{array}{cc}
\sflex_1 & \ \ \sflex_2 \\
\sflex_2 & \ \ -\sflex_1 
\end{array}
\right) \\ \nonumber
\sflex_{ij2}&=&-\frac{1}{2} \left(
\begin{array}{cc}
\sflex_2  & \ \ -\sflex_1 \\
-\sflex_1  & \ \ -\sflex_2
\end{array}
\right).
\label{eq:gdef}
\end{eqnarray}
In order to obtain a visual understanding of the flexion quantities,
we have used these forms for the $D_{ijk}$ matrix in terms of $\flex$
and $\sflex$ in order to calculate how a Gaussian image is transformed
by the various operations of weak lensing, according to equation
(\ref{eq:2ndorder}). The results are shown in Figure 1, which displays
the lensing operations in order of their spin properties. The Gaussian
galaxy is given a radius (standard deviation) of 1 arcsec; while the
convergence and shear imposed on the galaxy are realistic (10\% in
each case), the flexion is deliberately chosen to be extraordinarily
large for visualisation purposes (0.28 arcsec$^{-1}$, c.f. 0.04
arcsec$^{-1}$ intrinsic rms flexion on galaxies). We immediately see
the shapes induced by flexion: the first flexion leads to a
(vectorial, spin-1) skewness, while the second flexion leads to a
three-fold (spin-3) shape.

While the first flexion probes the local density via the gradient of
the shear field, the spin-3 second flexion probes the nonlocal part of
the gradient of the shear field. For example, consider a Schwarzschild
lens: the first flexion is by definition zero everywhere except at the
origin, as the gradient of the convergence is zero everywhere except
at the origin. However, there is certainly ``arciness'' generated by
such a lens; this is described by the second flexion. We will provide
explicit expressions for the first and second flexion generated by
simple mass distributions in Sections 4 and 5.

The series of lensing distortions can clearly be continued to
arbitrary order by taking permutations of additional spin-raising and
lowering derivatives. For instance the next order of distortion can be
decomposed into three fields; a spin-4 field, $\de\de\de\de \psi$, a
spin-2 field, $\de^* \de \de \de \psi$, and a spin-0 field, $\de^*
\de^* \de\de \psi$. The $n^{th}$ order term can be decomposed into
Int$(1+n/2)$ independent spin fields with spins $s=n$, $n-2$, $n-4$,
$\cdots$, $0$ if $n$ is even or $\cdots 1$ if odd. Consistency
relations similar to those for ${\cal F}$ and ${\cal G}$ can be found
for all the higher spin fields, which can also be used to estimate the
convergence field via Kaiser-Squires like relations (see Section 7).

However, in this paper we restrict ourselves to exploring the
possibilities given by the first and second flexion. We will now
proceed to calculate analytic expressions for both of the flexion
terms for simple lens models.

\section{Galaxy Halos: Circular Profiles}

In this section we present flexion predictions for galaxy-galaxy
lensing under the assumption of a circularly symmetric lens. This is
valid for a galaxy-galaxy lensing approach where we do not reorient lens
galaxies, resulting in a circularly averaged mean lens; in the
following section we will consider the impact of having elliptical
lenses. We consider a variety of different lens models, and show how
flexion can be used to constrain them.

\subsection{Flexion for the Singular Isothermal Sphere}

The approximately flat rotation curves observed in galaxies can be
most simply reproduced by a model density profile which scales as
$\rho \propto r^{-2}$.  Such a profile can be obtained by assuming a
constant velocity dispersion for the dark matter throughout the halo,
and so is known as the singular isothermal sphere (see e.g.
Binney \& Tremaine 1987).  The projected surface mass density of the
singular isothermal sphere (SIS) is

\begin{equation}\label{sigsis}
\Sigma(\xi)=\frac{\sigma^2_v}{2G\xi},
\end{equation}
where $\xi$ is the distance from the centre of the lens in the
projected lens plane and where $\sigma_v$ is the one-dimensional
velocity dispersion of `particles' within the gravitational potential of the
mass distribution, such as stars. The dimensionless surface mass density or
convergence is defined as $\kappa = \Sigma/\Sigma_c$, where $\Sigma_c$
(or the critical density) is defined as
\begin{equation}
\Sigma_c = \frac{c^2}{4 \pi G} \frac{D_s}{D_l D_{ls}},
\end{equation}
where $D_s$ and $D_l$ are the angular diameter distance to the source
and lens, respectively, and $D_{ls}$ is the angular diameter distance
between lens and source.  Thus for the case of the simple isothermal
sphere we have
\begin{equation}\label{ksis}
\kappa(\theta)=\frac{\theta_E}{2\theta},
\end{equation}
where $\theta=\xi/D_l$ is the angular distance from lens centre in the sky
plane and where $\theta_E$ is the Einstein deflection
angle, defined as
\begin{equation}
\theta_E=4\pi\left(\frac{\sigma_v}{c}\right)^2\frac{D_{ls}}{D_s}.
\end{equation}
The flexion, $\flex$, caused by the SIS at an angular vector
displacement, $\thetab$, from the lens centre on the sky plane is thus
simply
\begin{equation}\label{fsis}
{\bf {\cal F}} = -\left[\frac{\theta_E}{2\theta^2}\right]e^{i\phi},
\end{equation}
where $\phi$ is the position angle around the lens, and in this case
also gives the direction of the flexion.  The first flexion for this
profile is therefore circularly symmetric and (expressed as a vector)
directed radially inwards towards the lens centre, as would be
expected.

Similarly, the second flexion is:
\begin{equation}
{\cal G}=\frac{3\theta_E}{2\theta^2}e^{3i\phi}\ .
\end{equation}
This has a larger maximum amplitude than the first flexion for this
lens profile, fades off with the same power law index away from the
lens, and oscillates around the lens as a spin-3 quantity rather than
a spin-1 quantity.

\subsection{Flexion and Shear Derivatives}

Having considered the specific case of an isothermal sphere, we can
continue more generally with power law representations of the shear
around a lens:
\begin{equation}
\gamma=-A\theta^{-n}\ ,
\end{equation}
where $A$ is a constant, $n=1$ corresponds to an isothermal sphere,
$n=2$ corresponds to a point mass, and so on.  In particular, one can
ask whether one can better describe the arced nature of lensed objects
by the flexion we have defined, or the shear derivatives themselves.

In order to answer this question, for simplicity we rotate the system
such that the source lies along the $+x$ axis from the lens.  We then
consider what the second-order lensing amplitudes would be in a
``derivative-space,'' composed of the two non-zero shear derivatives:
\begin{equation}
\psi_D \equiv \left( \begin{array}{c} \gamma_{1,1}\\
\gamma_{2,2}\end{array} \right) = \left(
\begin{array}{c}
nA\theta^{1-n}\\
-2A\theta^{1-n}
\end{array}
\right).
\end{equation}
In ``flexion-space,'' where the components are the first and second flexions, the second-order lensing amplitudes are:
\begin{equation}
\psi_{\flex} \equiv \left(\begin{array}{c}
\flex \\
{\bf \cal G}
\end{array}\right)=\left(
\begin{array}{c}
(n-2)A \theta^{1-n}\\
(n+2)A \theta^{1-n}
\end{array}
\right).
\end{equation}
We wish to find out which is the most compact basis space. For any
given distribution, this will be the one for which only one eigenstate is
non-zero.  

\begin{figure}
\psfig{figure=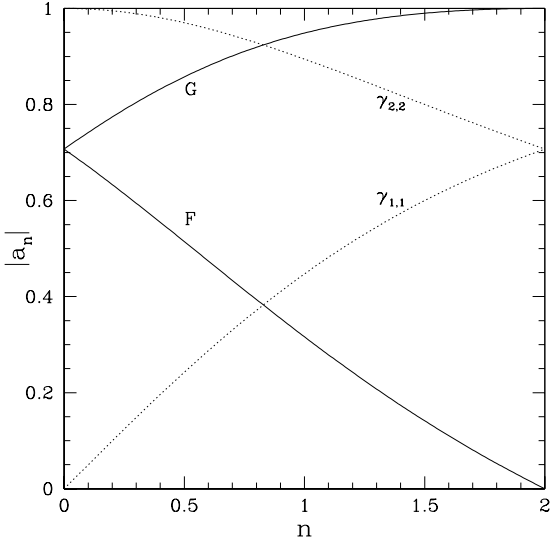,angle=0,height=3in}
\caption{Second order lensing amplitudes as a function of shear power
  law index. The solid line shows the amplitude of the flexion
  coefficients, and the dashed line shows the amplitude of the
  derivative coefficients.}
\label{fg:eigenspace}
\end{figure}

Figure \ref{fg:eigenspace} shows the amplitudes in each of these two
spaces as a function of the shear power law index. We see that, for
point sources, flexion space is the most compact approach; the signal
is a pure second flexion state. For a galaxy profile with $n \simeq
1$, both spaces are almost equally efficient in describing the second
order lensing.  Additionally, both the flexion and derivative
notations can be shown to produce 4 statistically independent terms,
which, taken over an ensemble of images will all have mean zero.
Moreover, {\it within} a representation, the standard deviations of the
two terms due to both intrinsic variation and photon noise will be
identical.

We conclude that flexion is an efficient means of describing third
order lensing. For point masses it is optimal; for SIS galaxies it is
as good as considering shear derivatives; and in addition the division
between local and non-local components which it exclusively affords is
very valuable. It also describes correctly the spin properties of the
lensing.

\subsection{Flexion for the Softened Isothermal Sphere}
\label{sissect}
The SIS mass distribution can be modified so as to remove one feature
which may not be a good physical description of dark matter
halos, the divergence of $\Sigma$ for $\theta \longrightarrow 0$. One
simple modification is to cut off the distribution at small distances
as follows:
\begin{equation}\label{kssis}
\kappa(\theta) = \frac{\theta_E}{2\sqrt{\theta^2+\theta_c^2}},
\end{equation}
where $\theta_c$ is a core radius within which the surface mass
density flattens off to a value $\kappa_0=\theta_E/2\theta_c$; it can
be seen that the projected mass distribution behaves like the SIS for
$\theta \gg \theta_c$. The flexion due to this distribution is
\begin{equation}\label{fssis}
{\cal F}= -\left[\frac{\theta_E}
{2(\theta^2+\theta^2_c)^{3/2}}\right] \theta e^{i\phi}\ .
\end{equation}
For $\theta \gg \theta_c$ the flexion is approximately equal to that
of the SIS. However, at small separations the flexion goes to zero, as
should be expected as the convergence is tending to a maximum.

The second flexion is more complicated:
\begin{equation}
{\cal G}=\frac{\theta_E}{2 \theta^3} \left(
-8\theta_c+\frac{3\theta^4+12\theta^2\theta_c^2+8\theta_c^4}{(\theta^2+
  \theta_c^2)^{3/2}} \right)e^{3i\phi}\ ,
\end{equation}
but may readily be fit to observed data, and can again be seen to
reduce to the SIS second flexion when  $\theta \gg \theta_c$ and goes to
zero at the centre of the lens.

\subsection{Flexion for the Navarro-Frenk-White (NFW) Density Profile}
\label{nfwsect}

Using N-body simulations, Navarro, Frenk \& White (1995, 1996, 1997)
have shown that the equilibrium density profiles of cold dark matter
(CDM) halos can be well fitted over two orders of magnitude in radius
by the formula

\begin{equation}\label{nfw}
\frac{\rho(x)}{\rho_{crit}(z)}=\frac{\Delta_c}{x(1+x)^2},
\end{equation}
where the radial coordinate $x$ is the radius in units of a scaling
radius $r_s$ such that $x \equiv r/r_s$, $\rho_{crit}(z)$ is the
critical density for closure at the epoch of the halo, and $\Delta_c$
is a dimensionless scaling density. This profile describes the
simulation halos accurately over a broad mass range $3 \times 10^{11}
< M_{200}/M_{\odot} < 3 \times 10^{15}$, $M_{200}$ being the total
mass of the halo contained within the sphere encompassing a mean
overdensity of 200 times the critical density $\rho_{crit}(z)$.  The
radius of this sphere, designated by $r_{200}$, is used to define a
second dimensionless scaling parameter for the NFW profile, namely the
concentration $c=r_{200}/r_s$.  However, the details of the NFW
definitions have been implemented in several ways in the literature;
Appendix A presents further discussion of the various definitions.

A procedure for finding values of $\Delta_c$ and $c$ which agree with
the numerical simulations is detailed by Navarro et al. (Appendix,
1997): the parameters are somewhat complicated functions of the halo
redshift and $M_{200}$, along with the background cosmology.  A
routine ({\tt charden.f}) which carries out these calculations and outputs
values for these scaling parameters has been made available by Julio
Navarro at {\tt http://pinot.phys.uvic.ca/}\~{\tt jfn/charden}.

The NFW density profile implies the following form for the dimensionless
surface mass density (Bartelmann 1996):

\begin{equation}
\kappa(y)=2\kappa_s\frac{f(y)}{y^2-1},
\end{equation}
where we define $\kappa_s=\rho_{crit}(z)\Delta_c r_s/\Sigma_{crit}$
and $y\equiv \xi/r_s$, with $\xi$ defined as for equation (\ref{sigsis}). The
function $f(y)$ is given by
\begin{equation}\label{fnfw}
f(y)=\left\{ \begin{array}{ll}
1-\frac{2}{\sqrt{1-y^2}}\mathrm{arctanh}\sqrt{\frac{1-y}{1+y}} & y<1 \\
1-\frac{2}{\sqrt{y^2-1}}\arctan\sqrt{\frac{y-1}{y+1}} & y>1.
\end{array} \right.
\end{equation}
The flexion for the NFW density profile is then given by
\begin{equation}
\flex\equiv\nablab_{\theta}\kappa 
=\frac{\partial y}{\partial \theta}\nablab_y \kappa.
\end{equation}
Defining $\mathcal F \it _s \equiv \kappa_s D_l / r_s $ we
then have
\begin{equation}
\flex=-\frac{2 \mathcal F \it _s}{(y^2-1)^2}
\left[ 2yf(y) - h(y) \right]
e^{i \phi}
\end{equation} 
with $y=\theta D_l/r_s = \theta / \theta_s$, and where, from
equation (\ref{fnfw}),
\begin{equation}
h(y)= \left\{\begin{array}{ll}
\frac{2y}{\sqrt{1-y^2}}\mathrm{arctanh}\sqrt{\frac{1-y}{1+y}}-\frac{1}{y} & y<1 \\
\frac{2y}{\sqrt{y^2-1}}\arctan
\sqrt{\frac{y-1}{y+1}}-\frac{1}{y} & y>1.
\end{array} \right.
\end{equation}
The analytical form of the second flexion can also be found,
using the fact that for axially symmetric projected mass profiles the
magnitude of the shear can be calculated from
$|\gamma(\theta)|=\bar{\kappa}(\theta)-\kappa(\theta)$, where
$\bar{\kappa}(\theta)$ is the mean surface mass density within a
circle of radius $\theta$ from the lens centre (see e.g. Bartelmann \&
Schneider 2001). Wright \& Brainerd
(2000) used this method to find an expression for the magnitude of shear
due to an NFW halo, and their result can be used to find the derivatives of
shear $\gamma_{1,1}$, $\gamma_{1,2}$ etc. Combining these derivatives
as directed by equation (\ref{eqn:fg}) we see that the second flexion
takes the form
\begin{equation}
\sflex = 2\flex_s\left[\frac{8}{y^3}\ln{\frac{y}{2}}+
\frac{\left(\frac{3}{y}(1-2y^2)+g(y)\right)}{(y^2-1)^2}
\right]e^{3i\phi},
\end{equation}
where
\begin{equation}
g(y)=\left\{ \begin{array}{ll}
\left(\frac{8}{y^3}-\frac{20}{y}+15y\right)
\frac{2}{\sqrt{1-y^2}}\atanh\sqrt{\frac{1-y}{1+y}} & y<1 \\
\left(\frac{8}{y^3}-\frac{20}{y}+15y\right)
\frac{2}{\sqrt{y^2-1}}\arctan\sqrt{\frac{y-1}{y+1}} & y>1.
\end{array} \right.
\end{equation}
To illustrate these results, we calculate the first and second flexion
signals we might expect to measure for a galaxy-sized halo with an NFW
profile.  We choose a lens redshift $z_l =0.35$ and the halo
$M_{200}=1\times 10^{12} h^{-1}M_{\odot}$, this lens redshift being
the median of the lens galaxy sample used by Hoekstra et al. (2004),
and the mass having been found to be roughly typical for galaxy halos
in weak lensing analyses by Brainerd et al. (1996) and Hoekstra et
al. (2004).  We also choose $D_{ls}/D_s = 0.5$ (corresponding to a
source redshift of $z_s \approx 0.8$) and model the lensing within a
standard, flat $\Lambda$CDM cosmology, setting the present-day matter
density parameter $\Omega_{m,0}=0.3$, $\Omega_\Lambda = 0.7$, the
Hubble parameter $h=0.72$ and $\sigma_8 = 0.8$.

\begin{figure}
\psfig{figure=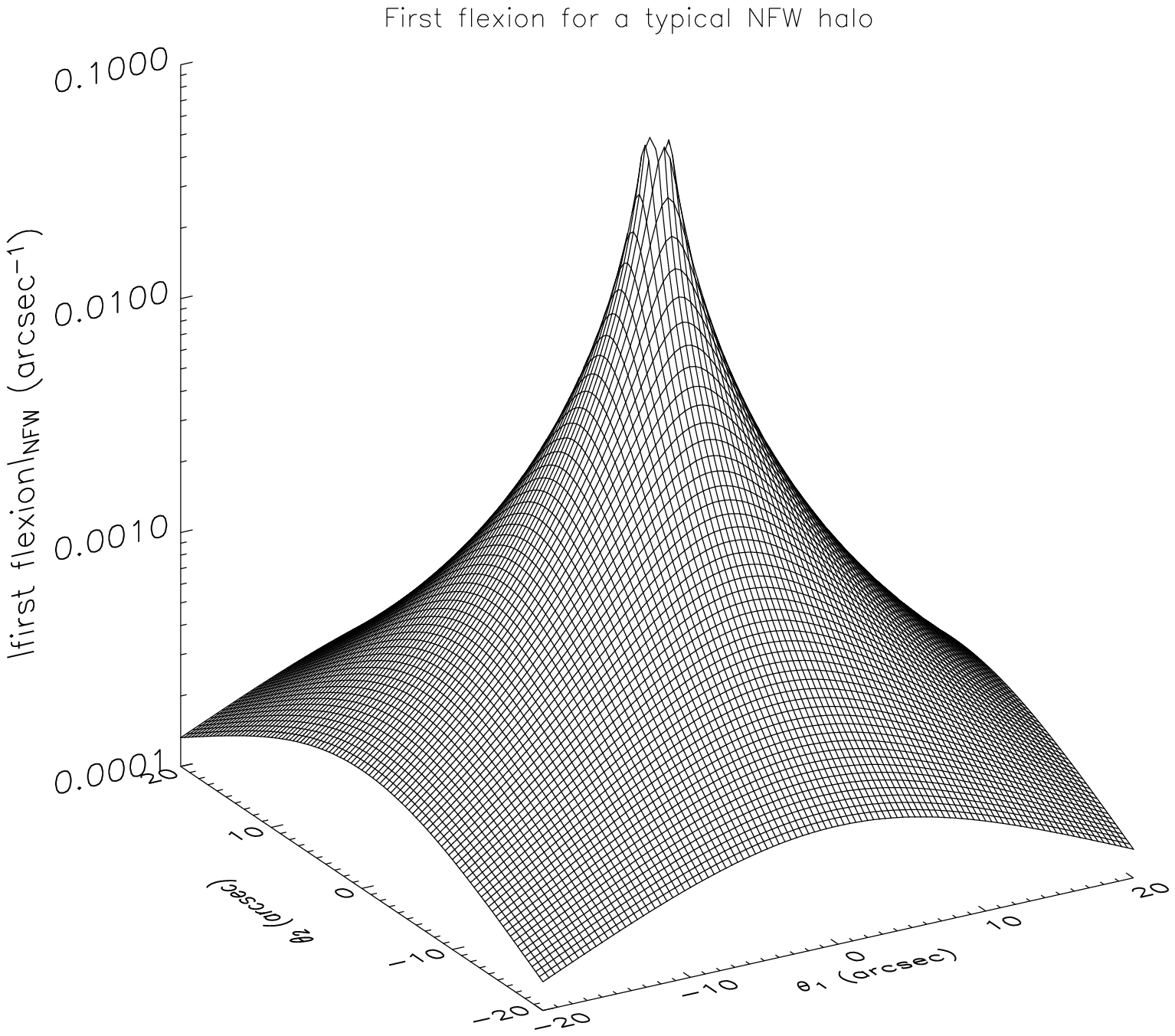,height=8cm,angle=0}\label{nfwfig}
\psfig{figure=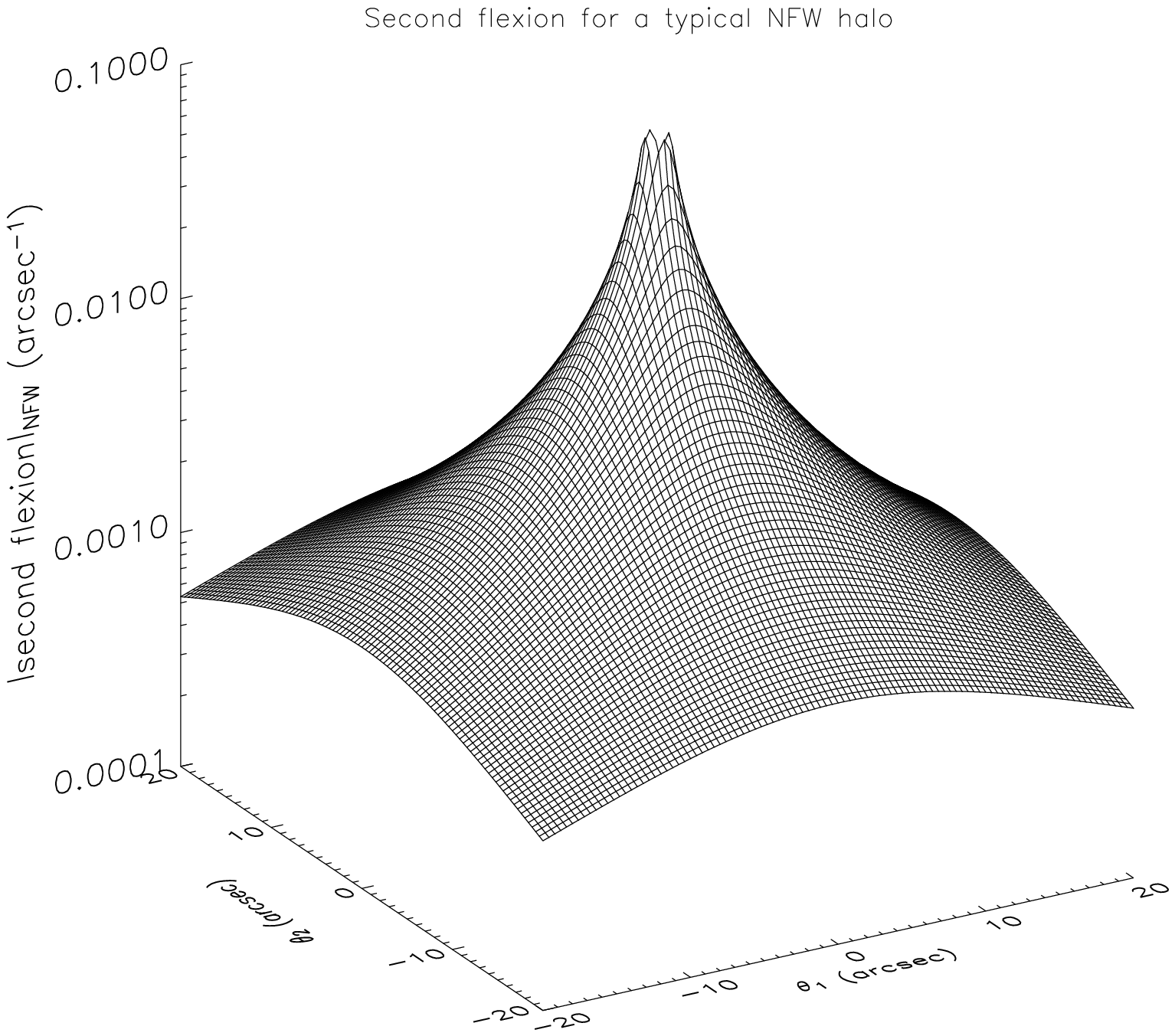,height=8cm,angle=0}\label{nfwfig}
\caption{Top: Logarithmic surface plot of the magnitude of first
flexion due to an NFW halo of $M_{200} = 1\times 10^{12}
h^{-1}M_{\odot}$ at redshift $z_{lens}=0.35$. Bottom: Logarithmic
surface plot of the magnitude of second flexion for the same halo.}
\end{figure}
Using these values and Navarro's {\tt charden.f} we find a concentration of
$c=7.20$ and a corresponding dimensionless characteristic density
$\Delta_c=2.03\times 10^4$. These values for the NFW parameters are
again in good agreement with those found by Hoekstra et al. (2004) who
measured $\Delta_c=2.4^{+1.4}_{-0.8}\times 10^4$ as the best fit to
their sample of $\sim 10^5$ lenses. The resulting flexion profiles are
shown in Figure 3; both flexion signals reaches a 1\% effect on a $\sim4''$
scale. We will now compare these flexion profiles with those resulting
from the SIS density profile, and will then discuss the measurability
of this signal with realistic survey models.

\subsection{A Comparison of the NFW and SIS Flexion Results}

Here we use the results from Sections \ref{sissect} and \ref{nfwsect}
to compare the flexion we would expect to measure for typical
galaxy-galaxy lensing observations, for the SIS and NFW cases.

\begin{figure}
\psfig{figure=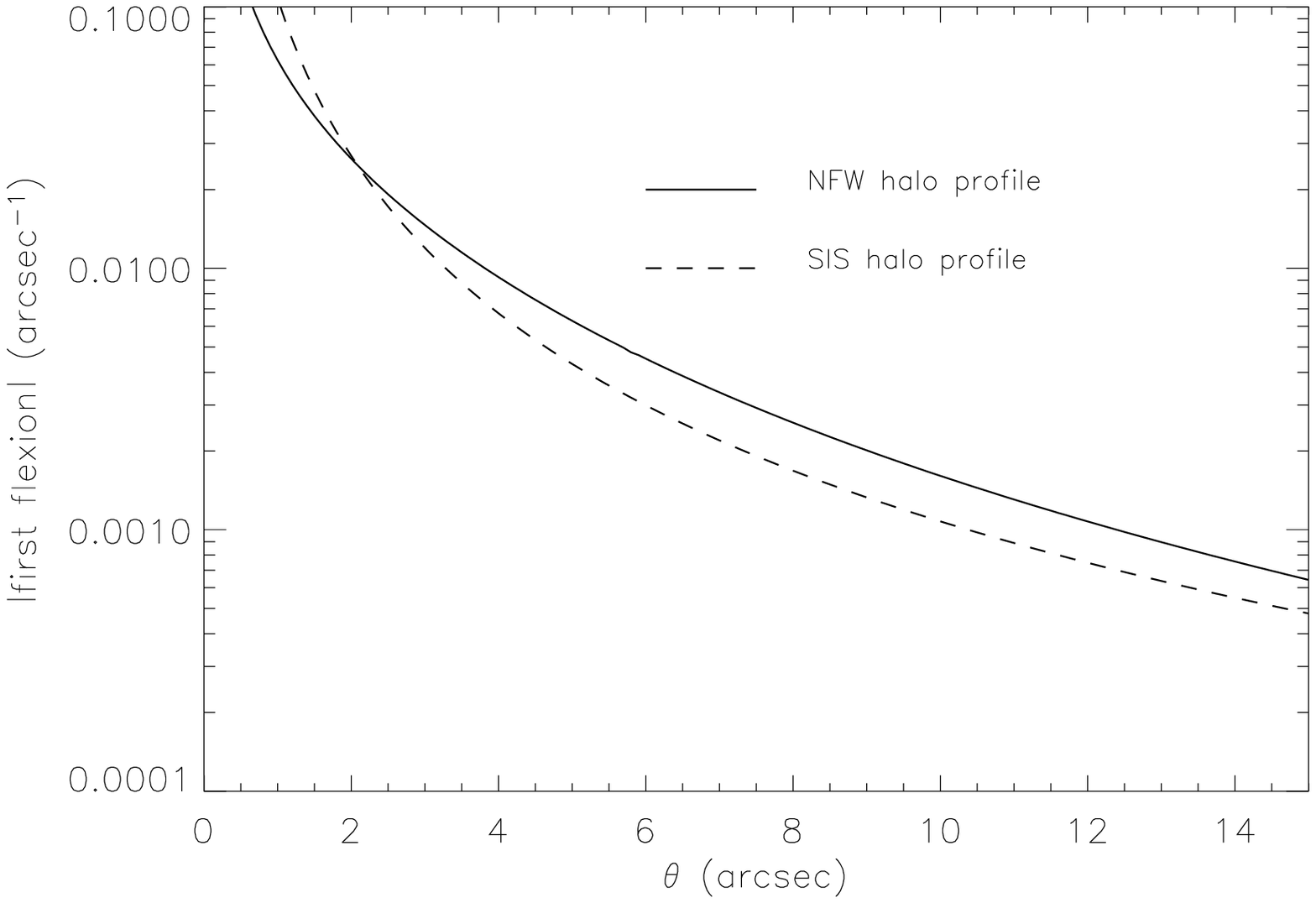,height=6cm,angle=0}\label{lognfwfig}
\psfig{figure=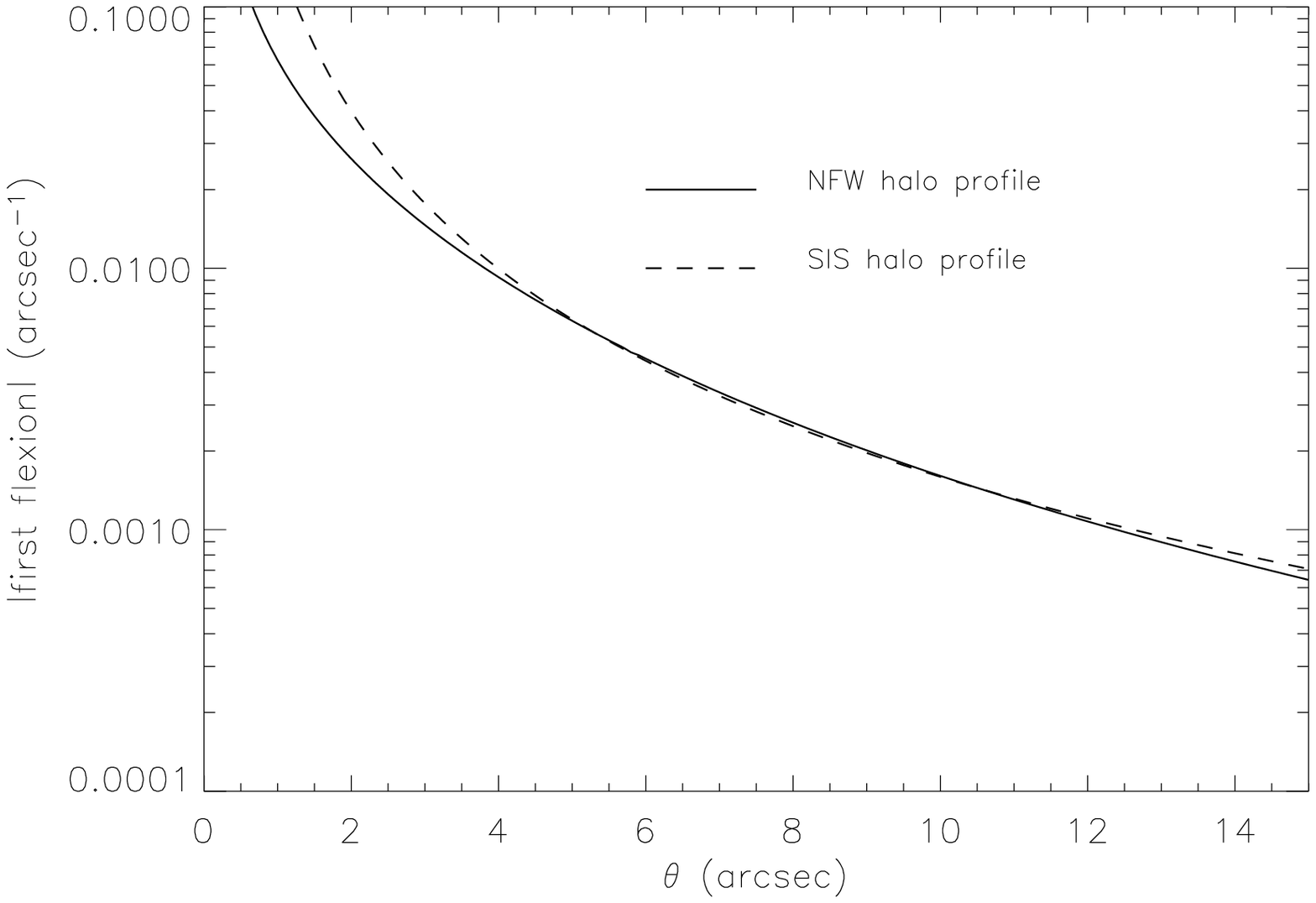,height=6cm,angle=0}
\psfig{figure=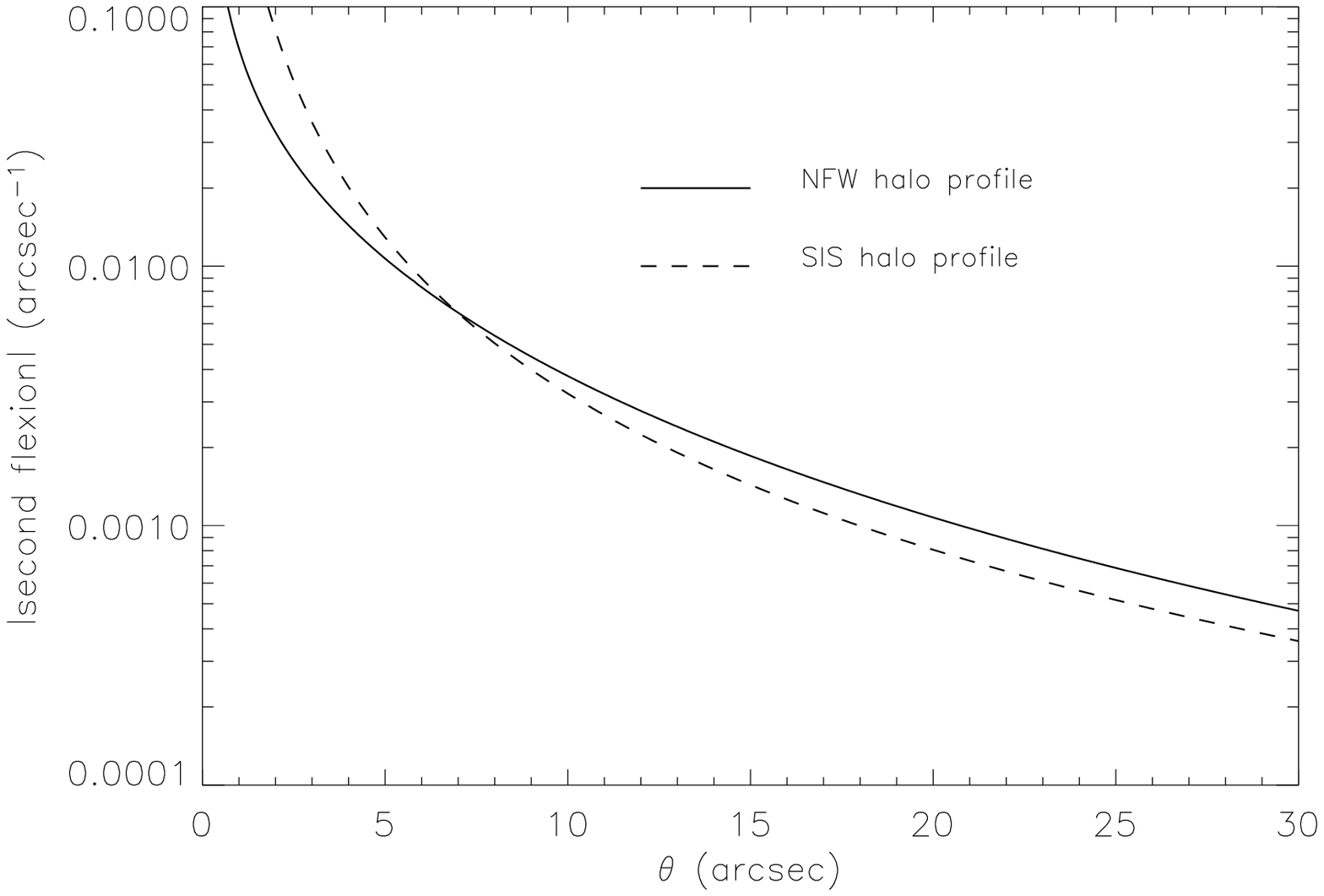,height=6cm,angle=0}
\caption{Top: Comparison of the magnitude of first flexion due to an NFW
and an SIS halo of $M_{200}=1\times 10^{12} h^{-1}M_{\odot}$ at
redshift $z_{lens}=0.35$. Middle: A similar $\flex$ comparison but
this time the SIS halo has $M_{200}=1.8 \times 10^{12}
h^{-1}M_{\odot}$. Bottom: The magnitude of $\sflex$ for an NFW and an
SIS halo of $M_{200} = 1\times 10^{12}h^{-1}M_{\odot}$, where the
doubling in scale of the angular separation axis highlights the larger
range and amplitude of the second flexion.}
\end{figure}

The SIS scaling is very straightforward in comparison to that of the
NFW halo; the Einstein radius for the SIS lens is given in terms 
of $M_{200}$ and the halo redshift $z_l$ as
\begin{equation}
\theta_E = \frac{2 \pi G}{c^2} \frac{D_{ls}}{D_s} \left(\frac{800 \pi
\rho_{crit}(z_l)}{3}\right)^{1/3} M^{2/3}_{200}.
\end{equation}
For this comparison we use the same values for $z_l$, $M_{200}$ and
the cosmological parameters as were used in Section \ref{nfwsect},
giving an Einstein radius for the SIS halo of $\theta_E = 0.215$ arcsec.

The predicted magnitudes of $\flex_{\rm NFW}$, $\sflex_{\rm NFW}$,
$\flex_{\rm SIS}$ and $\sflex_{\rm SIS}$, as a function of angular
separation from the lensing halo on the sky, are shown in Figure 4. As
could be expected the profiles show a good deal of similarity, but it
is apparent that both the first and second flexions due to the SIS
profile are stronger than those due to the NFW at very small
separations. Since one of the important features of the NFW profile is
that the density in the extreme interior of the halo varies as
$\propto r^{-1}$ compared to the steeper $\propto r^{-2}$ for the SIS,
this is not a surprising result.

It can be seen by comparing the lower plot of Figure 4, for which the
$\theta$ axis is doubled in scale, with the upper plot, that
$\sflex_{\rm NFW}$ is both stronger and longer range than $\flex_{\rm NFW}$.
Interestingly, we also note that the angular separation at which the
SIS halo flexion exceeds that for the NFW halo is larger by about 5
arcsec for second flexion in relation to the first flexion. These two
effects are a consequence of the non-locality of $\sflex$ as a lensing
measurement when compared to the directly local $\nablab \kappa$
measurement given by $\flex$; for the NFW profile, $\sflex$ tends to be less
steep than $\flex$ at small $\theta$ and to die away less rapidly at
larger separations.

The middle plot of Figure 4 shows another feature of the
comparison between the two profiles: an SIS halo of $M_{200}=1.8
\times 10^{12} h^{-1} M_{\odot}$ is practically indistinguishable from
an NFW halo with $M_{200}=1\times 10^{12} h^{-1} M_{\odot}$ for first flexion
measurements over galaxy-galaxy separations greater than about 5
arcsec.  This is a very similar property to one found by Wright \&
Brainerd (2000) in a comparison of the \emph{shear} profiles of SIS and NFW
halos.  They
found that the assumption of an SIS halo profile produced systematic
overestimations (by factors of up to 1.5) of the mass of NFW
halos. Further work will be required to determine the dependence of
this effect upon $c$ for flexion measurements as Wright \& Brainerd
usefully did for the case of shear.

\subsection{Combined Shear and Flexion - Improving NFW Halo Parameter 
Constraints} Previous studies of galaxy-galaxy lensing which have
aimed to constrain values of halo parameters such as $M_{200}$ or $c$
for the NFW profile (see for example Brainerd et al. 1996; Hoekstra et
al. 2004, hereafter HYG04 in this section; Kleinheinrich et al. 2005)
have used measurements of shear exclusively. Recently Goldberg \&
Bacon (2005) have shown that in many lensing scenarios the
signal-to-noise ratio will be larger for the flexion than for the
shear at small (but still easily measurable) angular separations
between source and lens. It is therefore worthwhile considering
whether combining measurements of shear and flexion might improve
constraints for the halo parameters such as $c$ or $M_{200}$ derived
from measurements of shear alone.

In order to do this we construct a simplified but illustrative
model. We can generate mock data for a sample of lens and source
galaxies such as might be available using current or forthcoming
galaxy imaging surveys. We model lens halos as NFW profiles, and (as
in HYG04) we assume we can scale each lensing measurement in the
sample to a fiducial mass $M_{200}$ or corresponding rest-frame B-band
luminosity $L_B$ using an observationally motivated scaling relation
between the two, such as that proposed by Guzil \& Seljak (2002).

In order to estimate the confidence limits we might reasonably expect
from weak lensing measurements, we must consider the effect of
\emph{intrinsic} ellipticity and flexion of unlensed galaxies.  We use
values of $\gamma_{int} = 0.2$ and $\flex_{int} = 0.04$ for the
intrinsic shear and flexion in this model (c.f. the intrinsic flexion
measured by Goldberg \& Bacon 2005).  Redshift errors must also be
considered; we assume for this simulation that we have access to
photometric redshifts for each galaxy, with an uncertainty of $\Delta
z$ on each individual redshift measurement (with values assigned below
for broad-band and medium-band photometric redshift surveys).

\begin{figure}
\psfig{figure=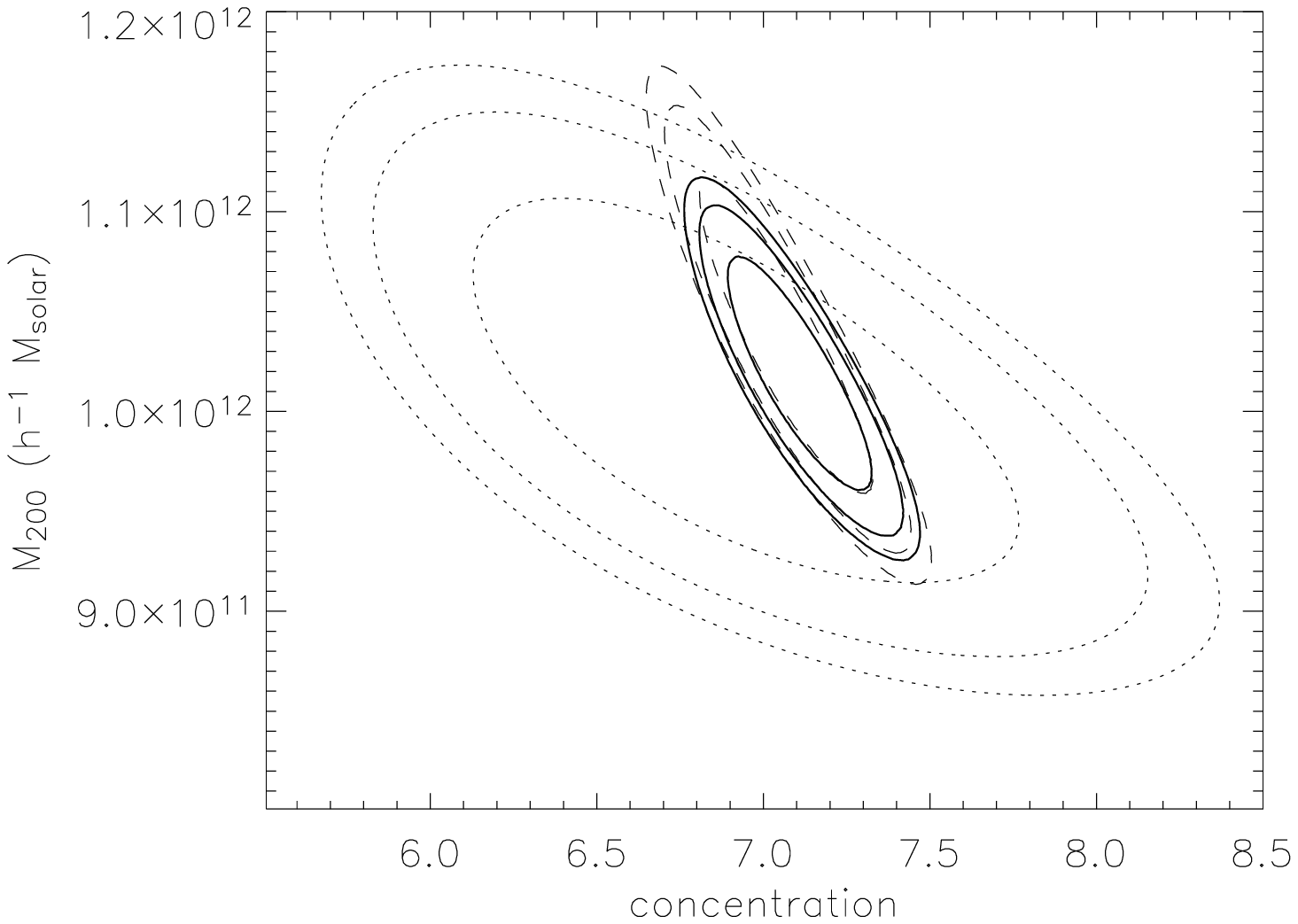,height=6.35cm,angle=0}\label{contours}
\psfig{figure=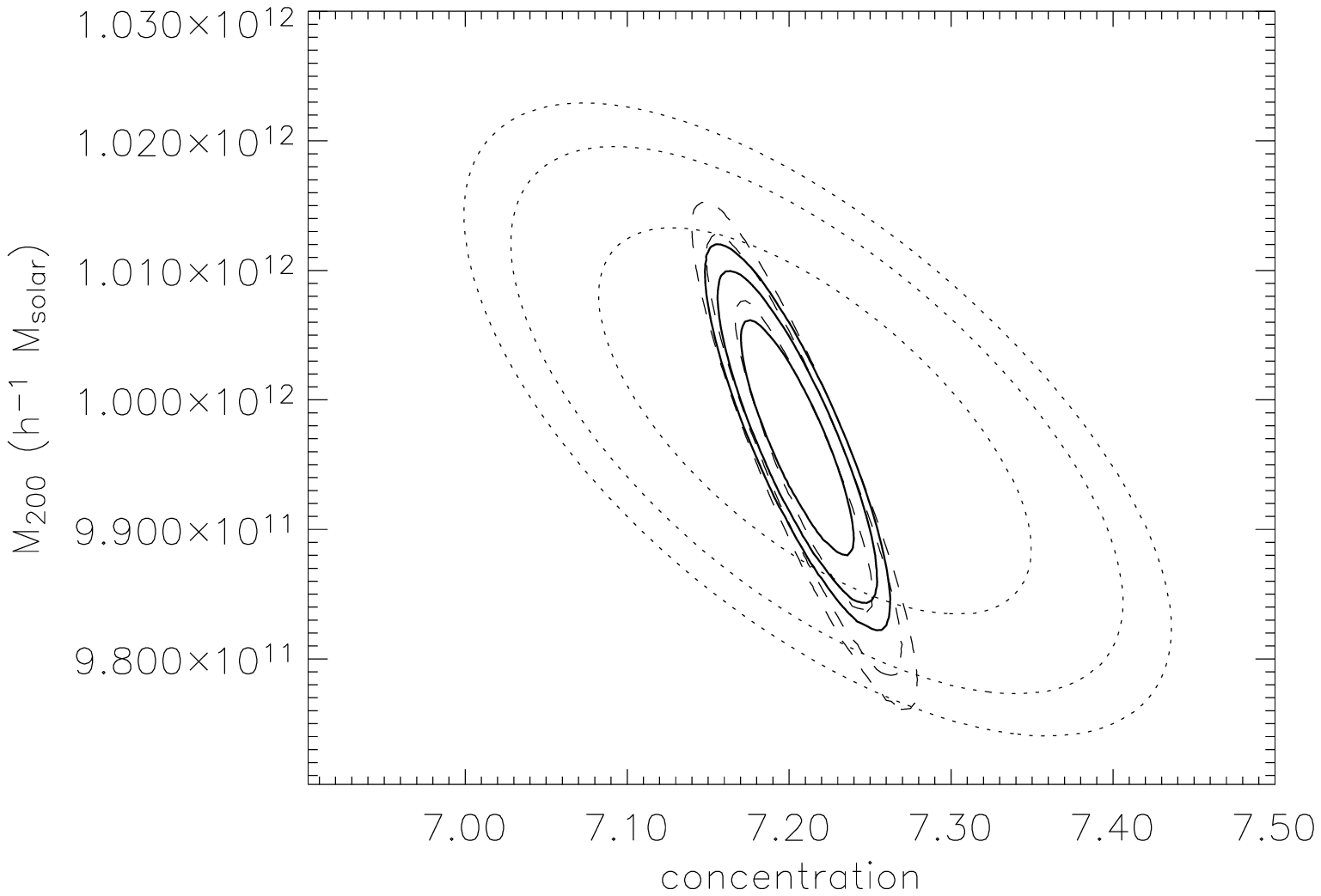,height=6.35cm,angle=0}
\psfig{figure=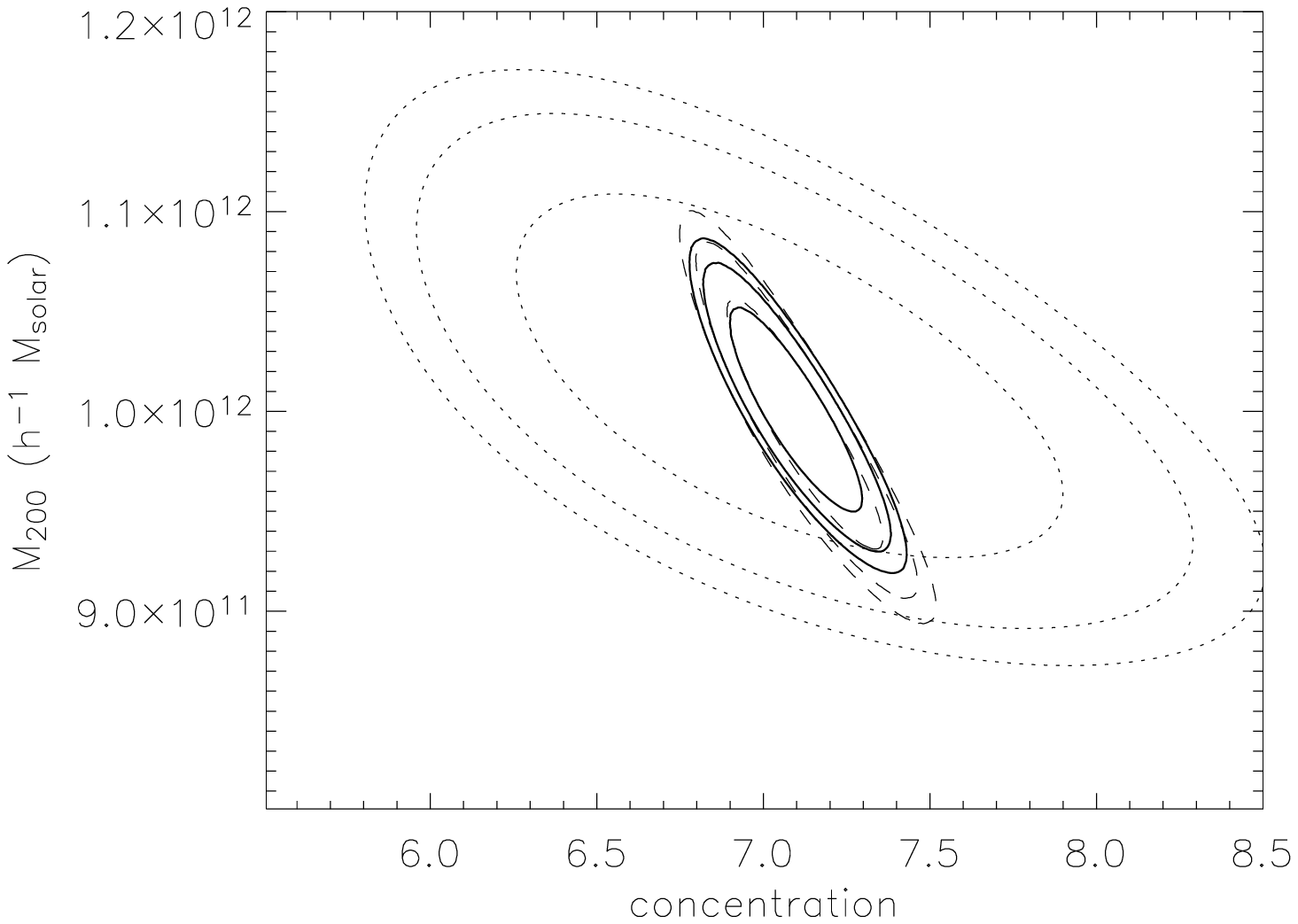,height=6.35cm,angle=0}
\caption{Estimated confidence limits on NFW halo parameters available
using (dotted line) shear measurements alone, (dashed line) flexion
measurements alone and (solid line) combined measurements of shear and
flexion. Top: a 42 sq deg ground-based survey such as that used by
Hoekstra et al. (2004). Middle: a 1700 sq deg ground-based
survey. Bottom: a 0.5 sq deg space based survey.}
\end{figure}

We note (e.g. Wright \& Brainerd 2000) that the strength of the shear
signal due to an NFW halo varies as $\gamma_{\rm NFW} \propto
D_lD_{ls}/D_s$, whereas we found in Section \ref{nfwsect} that the
strength of the flexion varies as $\flex_{\rm NFW} \propto D^2_l
D_{ls}/D_s$. We thus model the error on measurements of the shear and
flexion due to redshift uncertainties by calculating errors on $D_l
D_{ls}/D_s$ and $D_l^2D_{ls}/D_s$ by numerical integration of terms such as

\begin{equation}
\left\langle \left(\frac{D_l D_{ls}}{D_s}\right)^2 \right\rangle =
\int^\infty_0 \dif z_s' P(z_s'|z_s) \int^\infty_0 \dif z_l'
P(z_l'|z_l) \frac{D^2_{l'} D^2_{l's'} }{ D^2_{s'} } \label{bigD2}
\end{equation}
where $P(z_l'|z_l)$ and $P(z_s'|z_s)$ are the probability of measuring
a redshift $z_l'$ or $z_s'$ for a lens or source galaxy respectively,
given that its true redshift is $z_l$ or $z_s$. We model these
probability distributions as Gaussians with standard deviation $\Delta
z$, and assume a standard $\Lambda$CDM cosmology (as in Section
\ref{nfwsect}). We therefore estimate the fractional error in a single
measurement of shear and flexion due to redshift uncertainties (given
an underlying $z_l$ and $z_s$).  While the size of these fractional
errors depends upon each specific lens and source redshift, for the
purpose of this example we set them equal to the median
lens and source redshifts for each mock sample we consider.  Note that
while, if we had no redshift information, there would be a large
scatter in the signal caused by not knowing the geometry of the
lensing, this is drastically reduced with accurate photometric
redshifts and is assumed to be subdominant here.

For the fiducial virial halo mass we choose $M_{200} = 1\times 10^{12}
h^{-1}M_{\odot}$ (corresponding to a rest-frame L-band luminosity of
$L_B \approx 1.2\times 10^{10} h^{-2}L_{B,\odot}$ according to the
results of HYG04).  We choose to model confidence limits for two
ground-based surveys; one similar in size to that used by HYG04, and
one covering a substantially larger area of 1700 square degrees.
We also consider a deeper space-based imaging survey with far smaller
area of 0.5 square degrees.

The sample of galaxies used by HYG04 was taken from $R_c$ band imaging
of the the Red-Sequence Cluster Survey (Yee \& Gladders 2002) and
contained $N_l \sim 1.2\times 10^5$ lens galaxies and $N_s
\sim1.5\times10^6$ source galaxies over a sky area of 42 sq deg. This
corresponds to sky number densities of $n_l\approx 0.8$ arcmin$^{-2}$
for the lenses and $n_s \approx 10$ arcmin$^{-2}$ for the source
galaxies.  For the larger ground-based survey we assume the same
depth, but increase the survey area to 1700 sq deg.  We assume a
redshift uncertainty of $\Delta z = 0.1$ for each galaxy in either
sample, and use the median lens and source redshifts found by HYG04 of
$z_l=0.35$ and $z_s=0.53$ for both ground-based mock datasets. We set
the underlying NFW lens halo concentration to $c=7.20$ as in Section
\ref{nfwsect}.

For the mock space-based dataset we set the survey area to $0.5$ sq
deg, with number densities of $n_l=10$ arcmin$^{-2}$ and $n_s=30$
arcmin$^{-2}$ due to the increased depth and quality of imaging
expected for space-based results. For the redshift uncertainties we
use a value of $\Delta z = 0.05$ (c.f. Bacon et al. 2004 for the
COMBO-17 photometric redshift survey in relation to weak lensing; Wolf
et al 2001), and set $z_l=0.5$ and $z_s=1.0$. Following the
predictions of Navarro et al. (1997) we model each lens halo as having
a slightly smaller concentration of $c=7.02$ at this deeper redshift.

We then generate a set of mock results for the tangential shear and
radial flexion, averaged over annuli around the lensing galaxies (at
increasing angular separations between lens and source) for the whole
ensemble of galaxies in any given survey. These mock results are made
by taking the theoretical (NFW) prediction for the average shear or
flexion over each annulus of angular separation and offsetting it by a
Gaussian random deviate scaled to the estimated overall error for that
bin.

We combine the error due to redshift errors and the intrinsic signal
for a single measurement, multiplied by a factor of $1/\sqrt{N_{bin}}$
where $N_{bin}$ is the number of lens-source pairs within the annulus
over which we are averaging our lensing measurements. 

All that remains is to choose at what angular separations to impose
the divides between annuli for averaging shear and flexion
measurements. Since flexion is at its most useful on small scales,
while shear signals remain strong at scales large enough for the
flexion to become noise dominated, we divide up the angular scales for
measurement according to a geometric binning scheme. We choose 10
annuli such that the centre of the $i$th annulus lies at an angular
radius
\begin{equation}
r_i = a f^{(i-1)}
\end{equation}
where $a=2$ arcsec and the geometric factor $f = 1.5$.  In this way we
describe annuli which usefully cover both small (down to 2 arcsec) and
larger (up to 77 arcsec) scales of angular separation.

The resulting 68\%, 90\% and 95\%, 2-parameter confidence intervals for
NFW parameters from a maximum likelihood analysis of the three mock
datasets generated using this simple model can be seen in Figure 5; it
is immediately apparent that measurements of flexion may have a lot to
offer galaxy-galaxy lensing studies.  It is especially interesting to
note that the confidence-contours derived from measurements of shear
and flexion appear to be oriented at different angles in the plane,
allowing the two measurables to significantly complement each other.
This should perhaps not come as a surprise; whereas shear is a measure
related to the projected mass density $\kappa$, the first flexion
directly probes the local gradient of $\kappa$, or in this case the
slope of the halo profile. We should expect therefore that flexion has
the potential to significantly improve constraints on the halo
concentration $c$.

It is reassuring to note from Figure 5 that the size of the 68\%
confidence interval we derive on the fiducial $M_{200}$ for the
HYG04-like survey is in good agreement with the mass constraints found
by those authors for galaxies scaled to a (slightly smaller) fiducial
$L_B = 10^{10} h^{-2}L_{B,\odot}$, namely $M_{200} = (8.4 \pm 0.7 \pm
0.4) \times 10^{11} h^{-1} M_{\odot}$. The second error estimate
corresponds to a systematic uncertainty due to the fact that HYG04 had
no actual measured redshift information from the Red-Sequence Cluster
Survey (see HYG04 for details); we note that even despite this fact,
their errors due to intrinsic galaxy ellipticity dominate over redshift
uncertainties in their investigation of galaxy-galaxy shear, and will
therefore be even less dominant for surveys with measured redshifts.

\section{Galaxy Halos: Elliptical Profiles}

We now discuss the more general prospect of using flexion to measure
the ellipticity of lenses.  When describing elliptically flattened
halo mass distributions, it is often simplest to work with elliptical
lens potentials, $\psi(\thetab)$. Unfortunately these descriptions have
some severe limitations, most notably that they produce
dumbbell-shaped isodensity contours for large ellipticities and can
even produce negative surface-mass densities (see Kassiola \& Kovner
1993).

It is thus best to consider models where the isodensity contours of
the mass distribution are elliptical, despite the increased complexity
of the lens potential. The simplest generalisation of the softened
isothermal sphere to an elliptical density profile can be written

\begin{equation}\label{kellip}
\kappa(\theta_1,\theta_2)=\frac{\theta_E}{\displaystyle 2\sqrt{
\theta^2_c + \frac{\theta^2_1}{(1+\epsilon)^2} +
\frac{\theta^2_2}{(1-\epsilon)^2}}},
\end{equation}
where the major axis of the elliptical isodensity contours lie along
the $\theta_1$ axis in the sky plane, and the ellipticity $\epsilon$
is defined by the ratio of
minor-to-major axes ($b$ and $a$ respectively):
\begin{equation}
\frac{b}{a}=\frac{1-\epsilon}{1+\epsilon}.
\end{equation}
The flexion vector at $(\theta_1,\theta_2)$ in the sky plane is then

\begin{eqnarray}\label{fellip}
\flex&=&-\frac{\theta_E}{\displaystyle 2\left[
\theta^2_c + \frac{\theta^2_1}{(1+\epsilon)^2}+\frac{\theta^2_2}
{(1-\epsilon)^2}\right]^{3/2}} \nonumber \\
&\times&\left(\frac{\theta_1}
{(1+\epsilon)^2}+\frac{i\theta_2}{(1-\epsilon)^2}\right).
\end{eqnarray}
We note that interestingly, $\flex$ is no longer directed towards the
centre of the lens for all $(\theta_1,\theta_2)$; it will in fact be
centrally directed only when either $\theta_1$ or $\theta_2$ are equal
to zero.

It is simple to show that the flexion vector at a point
$(\theta_1,\theta_2)$ will be directed towards a point on the major
axis of the ellipse with coordinates $(a_{int},0)$ where

\begin{equation}\label{aint}
a_{int}=\left[1-\left(\frac{1-\epsilon}{1+\epsilon}\right)^2\right]
\theta_1=\left[1-\left(\frac{b}{a}\right)^2\right]\theta_1.
\end{equation}
Due to the $(b/a)^2$ term, even relatively modest ellipticities
in the density distribution cause $a_{int}$ to represent a significant
fraction of $\theta_1$. This tendency for the flexion vector to be
aimed at a point significantly off lens-centre can also be seen in
Figure 6, drawn for an axis ratio of $0.67$ which may be typical of
galaxy halos (see e.g. Hoekstra et al. 2004).  This implies that
measurements of the direction of flexion in galaxy-galaxy lensing may
be able to give good further constraints on the ellipticity of
dark-matter halos.

\begin{figure}
\psfig{figure=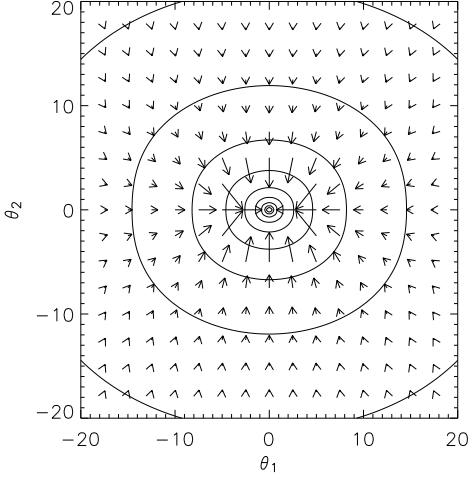,height=7cm,angle=0}\label{ellip}
\caption{Flexion vector field for an elliptical isothermal density
distribution with minor-to-major axis ratio of 0.67. Points in the
extreme interior of the diagram have been omitted for clarity and the
elliptical contours follow the logarithm of $|\flex|$.}
\end{figure}

In order to find the second flexion, we can rewrite this elliptical
isothermal profile (without softening) as follows.  We begin by defining
a radial term:

\begin{equation}
\rho\equiv \sqrt{\theta_1^2+f^2 \theta_2^2},
\end{equation}
where

\begin{equation}
f^2=(a/b)^2,
\end{equation}
with $a$ the semi-major axis and $b$ the semi-minor axis. The density
profile can then be defined as:

\begin{equation}
\kappa=\frac{A}{\rho}.
\end{equation}
For this distribution, the shear can be shown to have a very simple form:

\begin{eqnarray}
\gamma_1 & = -A\frac{\cos(2\phi)}{\rho} & =
  -A \frac{\theta_1^2-\theta_2^2}{\rho \theta^2}, \nn \gamma_2 & =
  -A\frac{\sin(2\phi)}{\rho} & = -A \frac{2\theta_1 \theta_2}{\rho
  \theta^2}.
\end{eqnarray}
We may compute the derivatives of these terms in a straightforward
way, and hence find the corresponding complex first and
second flexion:

\begin{equation}
\flex=\left(-\frac{A\theta_1}{\rho^3}\right)+i\left(-\frac{Af^2
  \theta_2}{\rho^3}\right)
\end{equation}
and
\begin{eqnarray}
\sflex=A\left(\frac{3\theta_1^5-\theta_1\theta_2^4-6\theta_1^3\theta_2^2-8f^2
  \theta_1\theta_2^4}{\rho^3 \theta^4}\right) + \nonumber \\ i A 
  \left(\frac{8\theta_1^4\theta_2+6\theta_1^2f^2\theta_2^3+f^2\theta_1^4t-3f^2\theta_2^5}{\rho^3
  \theta^4}\right).
\end{eqnarray}
The analysis becomes simpler if we only examine the angle-averaged
radial terms:
\begin{eqnarray}
{\cal F}_N=\langle -\exp(-i\phi){\cal F}\rangle = -\frac{A}{\rho
  \theta}\nn
{\cal G}_N=\langle -\exp(-3i\phi){\cal G}\rangle = \frac{3A}{\rho
  \theta}\ .
\end{eqnarray}
A means of measuring the ellipticity of the lens is to follow
Bartelmann \& Schneider (2001) and measure the quadrupole moment
of the flexion field over some aperture.  That is:
\begin{equation}
Q_{\{{\cal F},{\cal G}\}}=\int_0^{2\pi} d\phi \{ {\cal F}, {\cal G}\}\exp(2i\phi)\ .
\end{equation}
Despite the other advantages in simplicity of our mass model, the
evaluation of the the quadrupole moment here involves an elliptic integral.
However, for relatively small ellipticities, we can expand this out as a
series:
\begin{eqnarray}
Q_{\cal F}=-\frac{Ae}{8\theta^2}\simeq \frac{e}{8} \flex_N \nn
Q_{\cal G}=\frac{3Ae}{8\theta^2}\simeq \frac{e}{8} \sflex_N,
\end{eqnarray}
where $e$ is the lens ellipticity. Thus the lens ellipticity
measurement from flexion incurs an $e/8$ ``penalty'' compared to the
simple measurement of the flexion itself.  Taking a typical
ellipticity of $0.2$, the quadrupole estimate is $0.025$ times the S/N
of the flexion, and thus we need approximately 1600 times as many
pairs in order to measure the lens ellipticity effectively than to
measure the convergence field. Nevertheless, flexion can clearly
contribute to the question of the shape of dark matter halos around
galaxies.

This concludes our examination of galaxy-galaxy flexion prospects. We
will now turn to another area in which flexion can contribute
significantly to studies of the dark matter distribution: that of
mapping the dark matter density.

\section{Mass reconstruction and substructure}

In this section, we discuss how flexion can be used to reconstruct the
density field of matter in order to obtain a spatial map of the matter
distribution. This is clearly a valuable aspect of lensing, and is
already routinely achieved using weak shear. In addition, we can
obtain matter maps from flexion, which as we will see can
significantly improve the signal-to-noise of the density map. We will
first examine how to use flexion to obtain 2-D surface density maps of
matter; we will then examine how flexion can also be used for 3-D
mapping of density.

\subsection{2-D Mapping}

For 2-D mapping, we are able to generate maps of the projected matter
density (i.e. the convergence) from both $\flex$ and $\sflex$, following the
ideology of Kaiser and Squires (1993). Starting with $\flex$, we take the
Fourier transform of the relation $\flex_i=\partial_i \kappa$ to obtain

\begin{eqnarray}
\tilde{\flex_1}&=&-i k_1 \tilde{\kappa}({\bmath k}) \nn
\tilde{\flex_2}&=&-i k_2 \tilde{\kappa}({\bmath k}).
\end{eqnarray}
We can invert both of these terms to obtain an estimate for
$\tilde{\kappa}$. We add these two estimates in an optimal fashion,
parameterised by the variable $a$:

\begin{equation}
\tilde{\kappa}=\frac{i a \tilde{\flex_1}}{k_1}+\frac{i (1-a)\tilde{\flex_2}}{k_2}.
\end{equation}
In order to optimise the estimate, we take the mean square of this
equation, which in the absence of a lensing signal will have a value
determined by constant noise from intrinsic flexion. We then minimise with
respect to $a$, in order to find a measurement of the $\kappa$ field
with minimal noise. As a result we find the following inversion:

\begin{equation}
\tilde{\kappa} =
\frac{i k_1}{k_1^2+k_2^2}\tilde{\flex_1}+\frac{i k_2}{k_1^2+k_2^2}\tilde{\flex_2}.
\label{eqn:mapf}
\end{equation}
This gives us a prescription for finding the surface density of
matter: we measure the flexion field, take the Fourier transform,
calculate $\tilde{\kappa}$ according to this equation, and then take
the inverse Fourier transform to find $\kappa$.

We can perform the same calculation for the inversion from $\sflex$ to
$\kappa$. We note that the components of $\sflex$ can be written in terms
of the lensing potential, $\psi$ (c.f. equation \ref{eqn:fg}) as

\begin{figure}
\psfig{figure=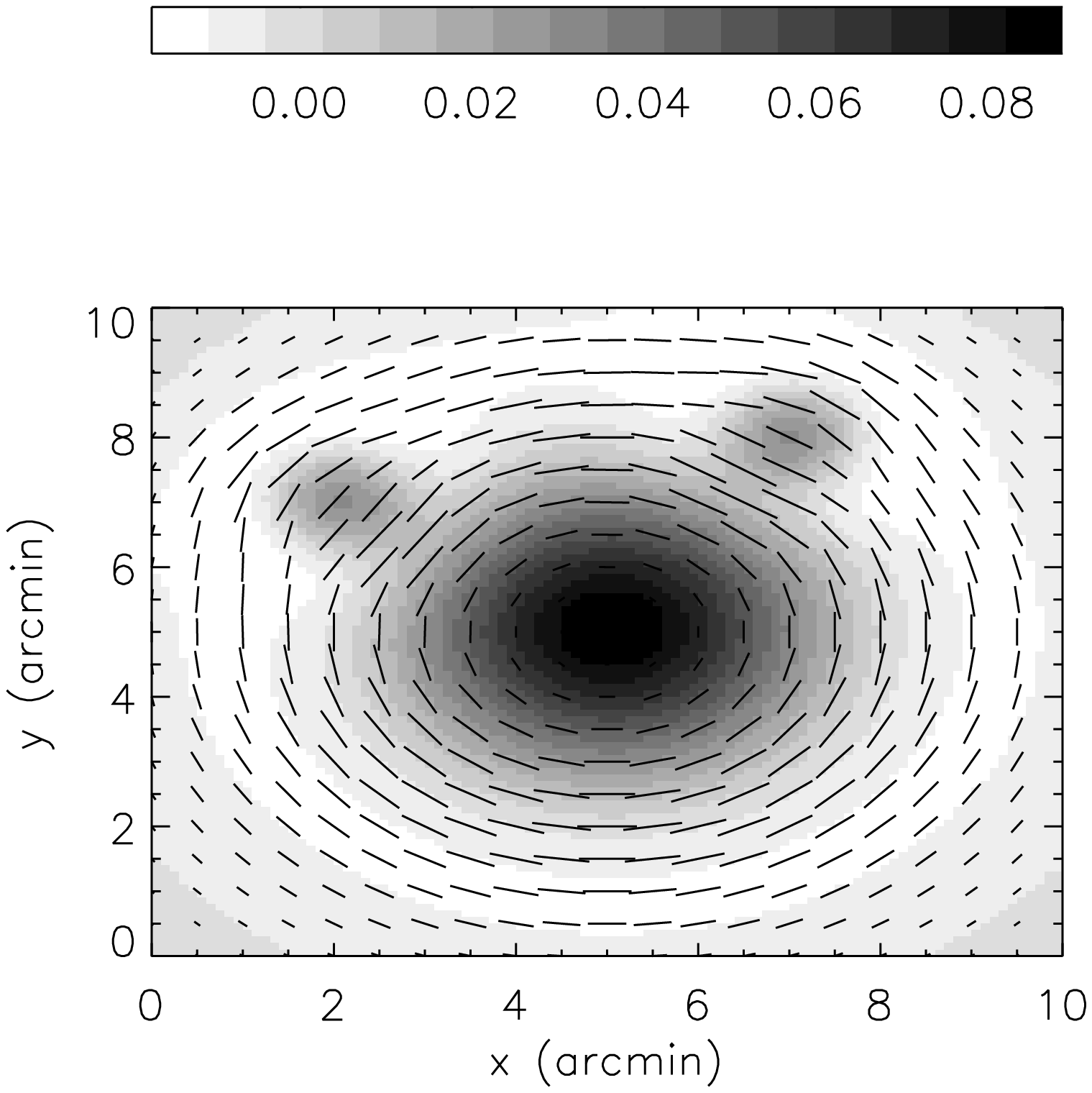,height=7cm,angle=0}
\psfig{figure=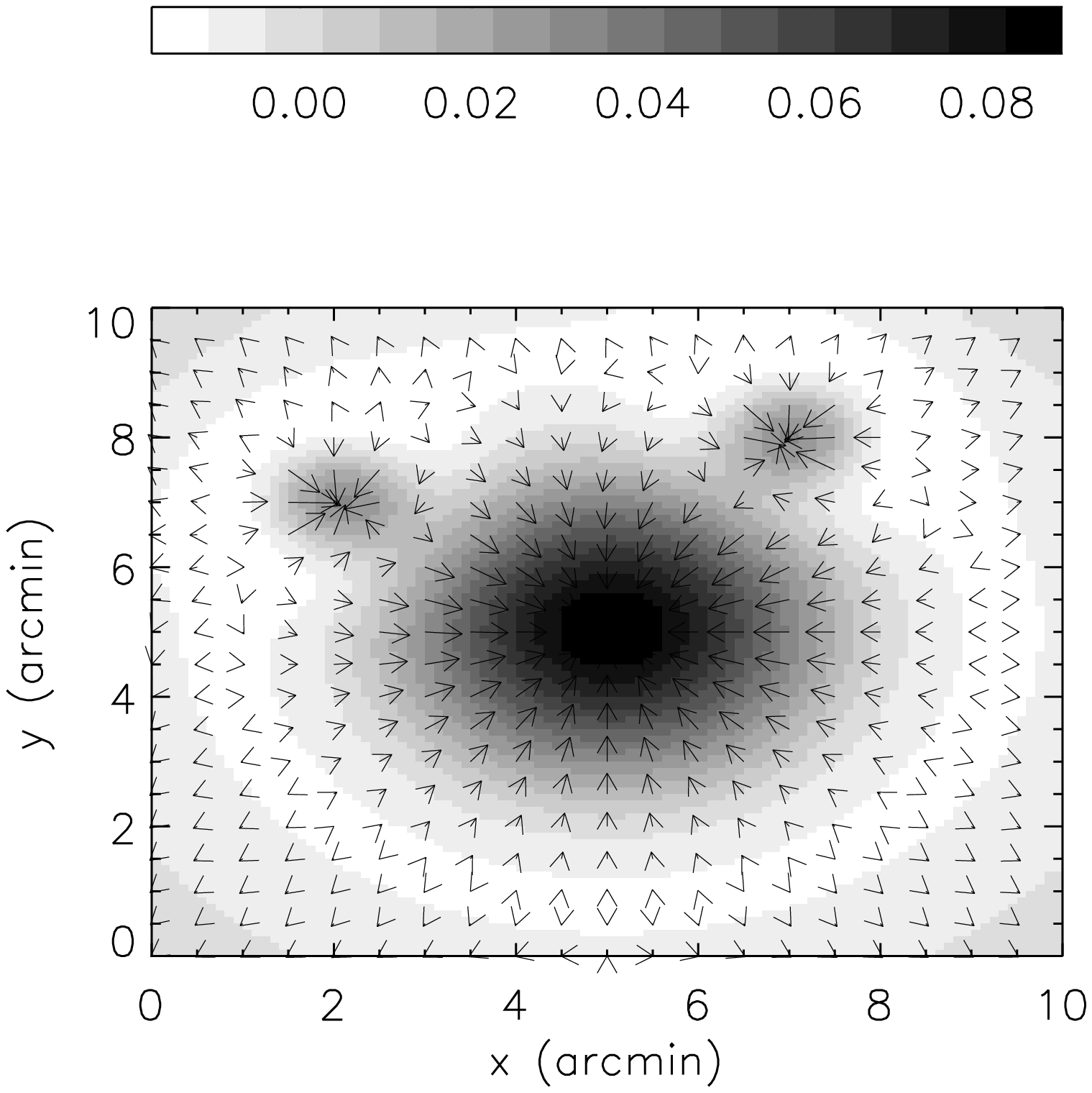,height=7cm,angle=0}
\psfig{figure=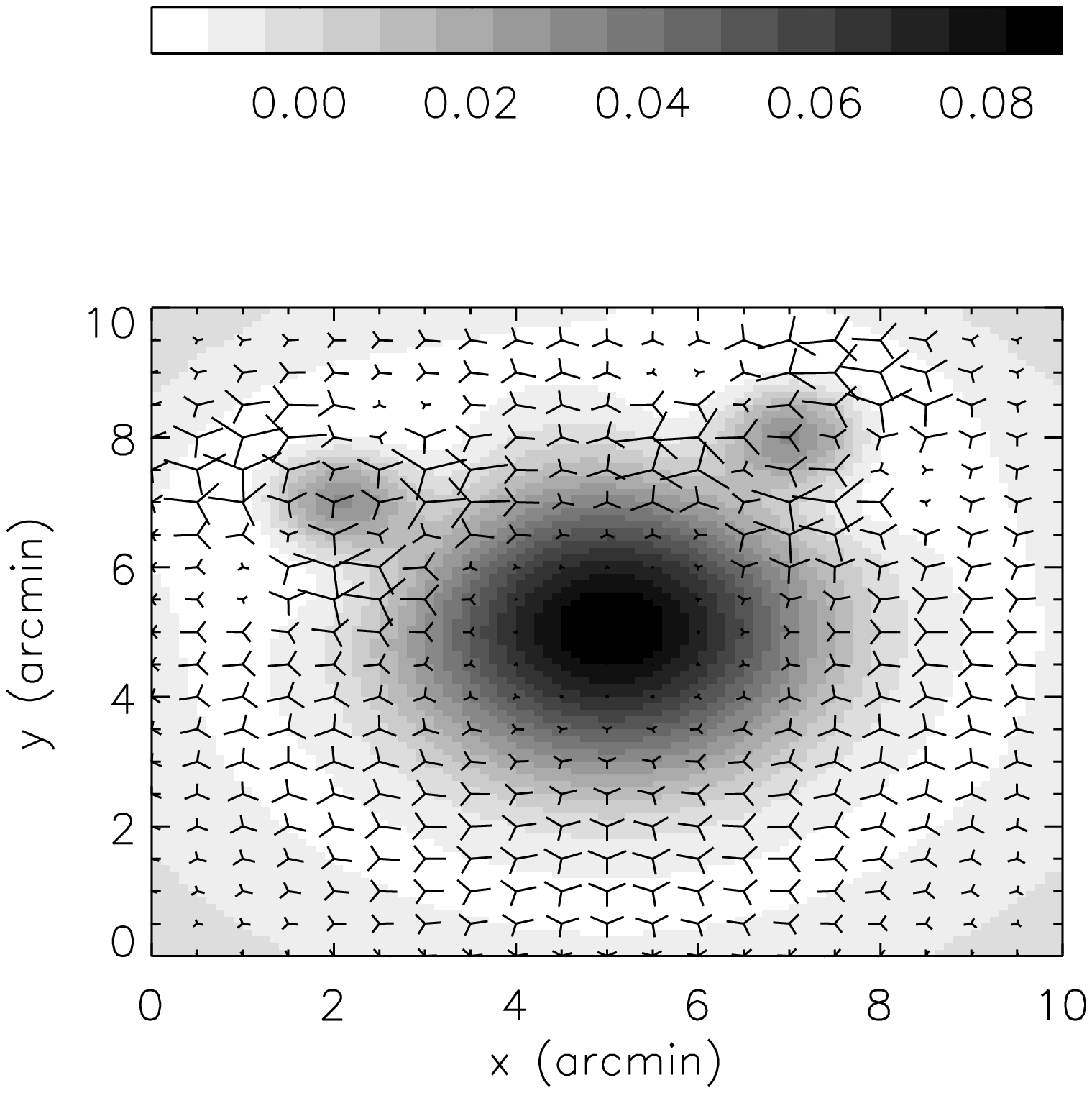,height=7cm,angle=0}
\caption{Shear (upper), flexion (middle) and second flexion (lower)
for simulated cluster; the cluster's convergence map is shown
underlying the other weak lensing fields. Note that shear does not
respond well to substructure, while the flexions respond very
significantly to these regions.}
\end{figure}

\begin{figure}
\psfig{figure=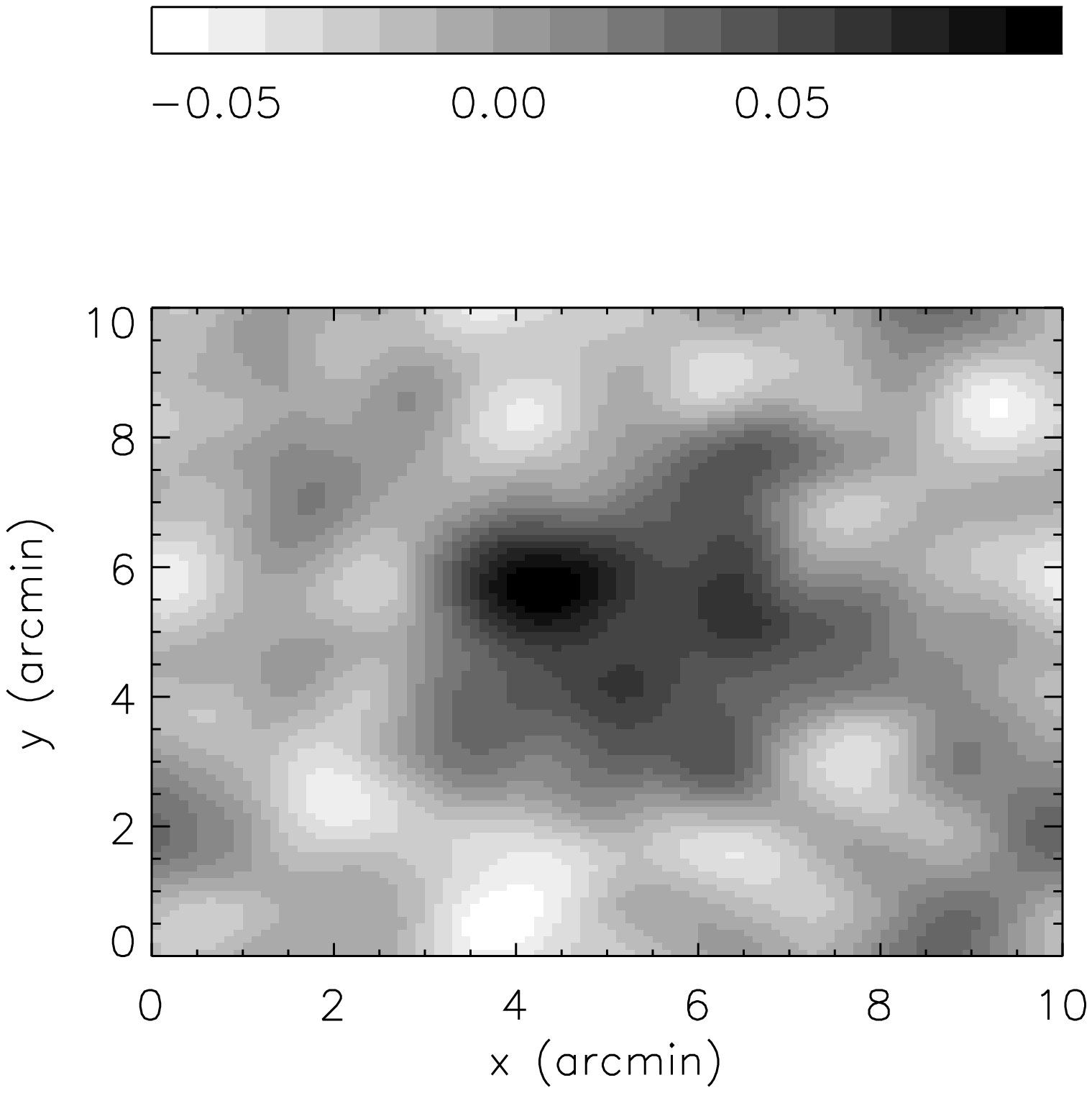,height=7cm,angle=0}
\psfig{figure=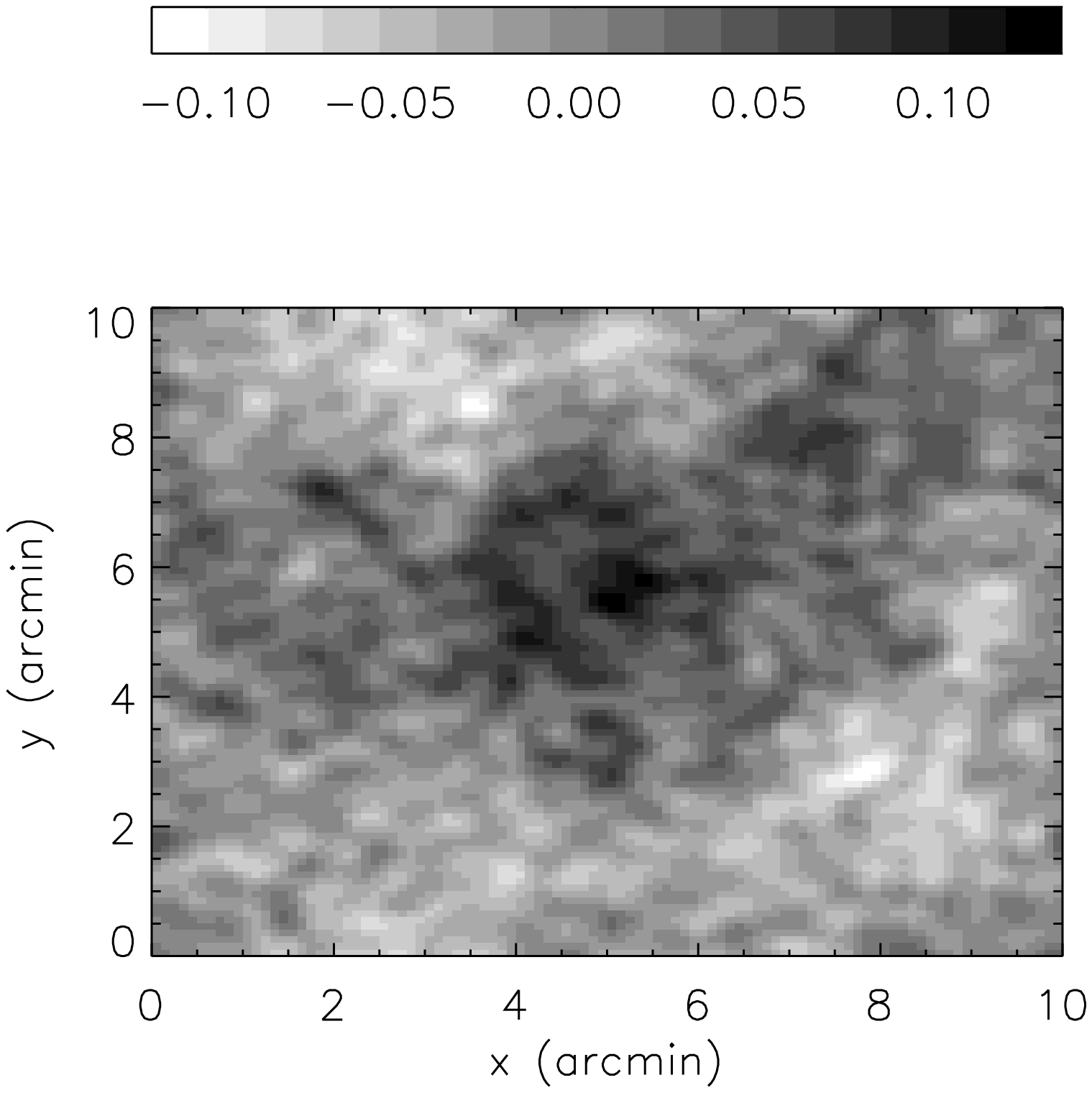,height=7cm,angle=0}
\psfig{figure=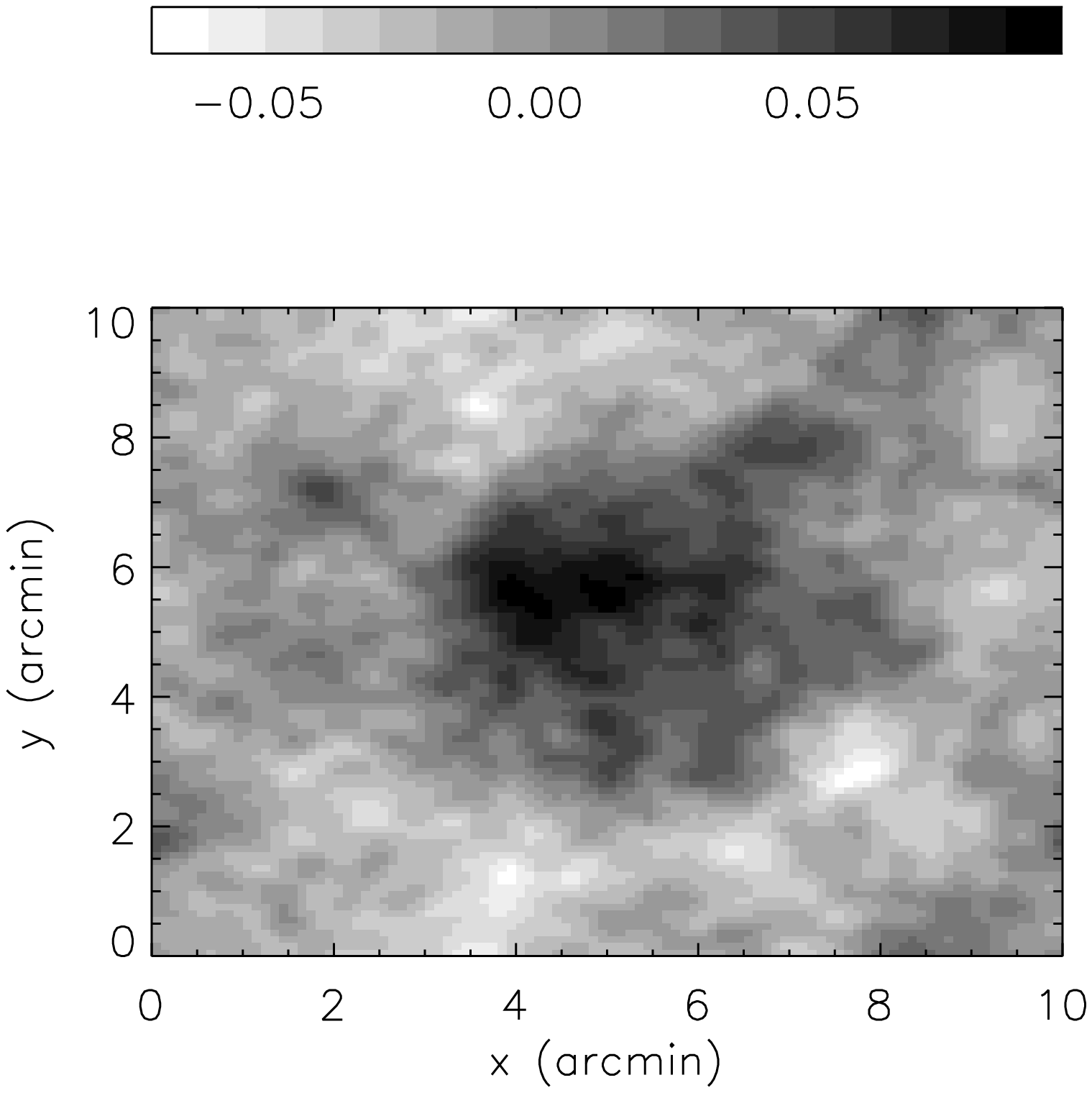,height=7cm,angle=0}
\caption{Recovered convergence maps from the shear alone (upper), the
two flexion fields (middle) and shear and flexion combined (lower) for
simulated cluster, with noise properties appropriate for a deep
space-based set of observations.}
\end{figure}

\begin{eqnarray}
\sflex_1 = (\partial_1^3-3\partial_1\partial_2^2)\psi \nn
\sflex_2 = (3\partial_1^2\partial_2-\partial_2^3) \psi.
\end{eqnarray}
Hence the Fourier transform

\begin{eqnarray}
\tilde{\sflex_1} &=& i (k_1^3-3 k_1 k_2^2) \tilde{\psi} \nn
\tilde{\sflex_2} &=& i (3k_1^2 k_2 - k_2^3)\tilde{\psi}.
\end{eqnarray}
Again, we add these estimates of $\tilde{\psi}$ in some optimal
fashion parameterised by $a$:

\begin{equation}
\tilde{\psi}= -\frac{i a\tilde{\sflex_1}}{k_1^3-3 k_1
k_2^2}-\frac{i (1-a)\tilde{\sflex_2}}{3k_1^2 k_2 - k_2^3}.
\end{equation}
Calculating the mean square of this field and minimising with respect
to $a$, we find that the optimal estimate of $\kappa$ is given by:

\begin{equation}
\tilde{\kappa}=i \frac{k_1^3-3k_1k_2^2}{(k_1^2+k_2^2)^2}\tilde{\sflex_1}+i
\frac{k_2^3-3k_1^2k_2}{(k_1^2+k_2^2)^2}\tilde{\sflex_2}.
\label{eqn:mapg}
\end{equation}
This provides us with the mass-mapping equations we have been
seeking. We can now obtain mass maps with independent noise for
$\gamma$, $\flex$ and $\sflex$, and combine these with minimum
variance weighting (with respect to noise) in order to obtain a best
mass map.

These mapping relations can be efficiently expressed and trivially
derived in the complex notation of Section 3 using equation (\ref{eqn:dfg}):

\ba
    (\kappa + i B)_{\cal F} &=& \de^{-2} \de^* {\cal F} ,\nn
    (\kappa + i B)_{\cal G} &=& \de^{-4} \de^* \de^*\de^* {\cal G}
\label{complexmap}
\ea
where the complex part is again seen to give us the B-field component
which can be used as a test of systematics. Comparing these two
derivations of the mapping equations, we see that (\ref{complexmap})
gives the solution in the case of no noise, while (\ref{eqn:mapf}) and
(\ref{eqn:mapg}) show that this is still optimal in the presence of
noise due to intrinsic flexion.

The mapping process is illustrated in Figures 7 and 8. Here we have
simulated a projected surface density for a toy cluster of galaxies,
using a Gaussian cluster gravitational potential profile with width
$\sigma=3'$ and mean $\kappa$ within this radius of $\kappa=0.06$. We
have laid down three substructure Gaussians containing 10\% of the
mass, with width $\sigma=1'$ (one at the centre of the cluster). The
associated shear and flexion fields shown in Figure 7 were calculated
directly from equations (\ref{eqn:gamma}) and (\ref{eqn:fg}). Note
from this figure that the shear does not respond significantly to the
small-scale structure, while flexion is most affected at these scales;
this is in line with our results for galaxy-galaxy flexion, and will
be explored more in the following section. We also note from the figure
that the first flexion responds locally to the density gradient,
whereas the second flexion responds non-locally while still giving
large signals near substructure.

Shot noise is added to these fields with $\sigma_\gamma=0.2$,
$\sigma_{\flex}=\sigma_{\sflex}=0.04$ and projected number density $n=60$ as
appropriate for a space-based survey such as GEMS (e.g. Rix et al 2004).

We have then used our inversion procedure (equations \ref{eqn:mapf}
and \ref{eqn:mapg}) together with the Kaiser-Squires inversion for
shear, to obtain maps of $\kappa$ from these fields, which are
displayed in Figure 8 together with a combined convergence map from
all fields added with minimum variance weighting. The shear field has
been smoothed with a Gaussian of radius 0.5' as it suffers from large
fluctuations on small scales, while the flexion is smoothed with
radius 0.1' as does not suffer from this problem. We note that the
surface density is reconstructed well from all three fields, with
maximum signal-to-noise of 3.6 for the shear reconstruction and 3.5
for the two flexion reconstructions combined. It is gratifying that
the signal-to-noise for the two approaches are so similar, and
strongly emphasises the value of measuring flexion as well as
shear. We also note that flexion does indeed measure the substructure
concentrations at the 1.4-2.6$\sigma$ level, whereas shear is not able
to detect these subhalos. Future lensing maps of density will
therefore benefit significantly from the inclusion of the flexion
signal, especially for the purpose of charting the substructure.

\subsection{3-D Mapping}

We will now briefly note how to extend this method in order to map the
density of matter in three dimensions with flexion, following the concepts
of Taylor (2001) and Bacon \& Taylor (2003). For this, we need to know
what gravitational flexion we would measure upon a galaxy at any 3-D
point in the Universe. We will see in the next section that the
effective flexion along a line of sight over cosmological distances is
given by

\begin{equation}
\flex=\frac{3H_0^2 \Omega_m^2}{2 c^2}\int_0^w dw' \frac{w'^2 (w-w')}{a(w')
w} \frac{\partial \delta}{\partial x}
\label{eqn:feff}
\end{equation}
where $H_0$ is the Hubble constant, $\Omega_m$ is the matter density
at the present epoch, $c$ is the speed of light, $w$ is comoving
distance, $a$ is the expansion factor, $\delta$ is the overdensity of
matter and $x$ is the transverse physical distance.

Now for a function $A(w)$ that can be written as the integral of a
function $B(w',w)$,

\begin{equation}
A(w)\equiv \int_0^w dw' B(w',w) ,
\end{equation}
we can write the rate of change of $A$ with respect to $w$ as

\begin{equation}
\frac{\partial A(w)}{\partial w}=\int_0^w dw' \frac{\partial
B(w,w')}{\partial w}+B(w,w) .
\label{eqn:ab}
\end{equation}
Now $\flex$ is in a suitable form for $A$, with $B$ given in equation
(\ref{eqn:feff}). We can therefore use equation (\ref{eqn:ab}) to
invert the integral for $\flex$, and find that the transverse gradient
of the matter overdensity, $\delta'$ can be calculated in terms of the
measured 3-D flexion:

\begin{equation}
\delta'(w)=\frac{2 c^2}{3 H_0^2 \Omega_m} \frac{a(w)}{w^2}
\frac{\partial^2}{\partial w^2} (w \flex).
\end{equation}
Thus we can obtain estimates of the density gradient along a line of
sight, if we have measurements of $\flex(w)$ along that line of sight,
improving signal-to-noise from 3-D maps measured using weak shear alone
(Taylor et al 2004).

\section{Cosmic Flexion}

We now turn our attention from dark matter mapping to the overall matter
distribution in the Universe. Can we use flexion to probe the
distribution of large-scale structure? In order to answer this
question, we carry out an analysis which is analogical to the
theory of cosmic shear; here, we are trying to calculate the `cosmic
flexion', the flexion correlation function whose signal originates
from the large-scale structure. In this section we will closely follow
the analysis of Bartelmann \& Schneider (2001).

\subsection{Flexion Power Spectrum}

If we are to find the flexion correlation function from large-scale
structure, then from the definition of flexion as the gradient of the
convergence, it is valuable to begin with the cosmological effective
convergence, given by Bartelmann \& Schneider (2001) as:

\begin{equation}
\kappa({\bf \theta},w)=\frac{1}{c^2}\int_0^w dw'
\frac{(w-w')w'}{w}\frac{\partial^2}{\partial x_i \partial x_i}
\Phi[w'{\bf \theta}, w']
\end{equation}
where ${\bf \theta}$ is the position on the sky, $w$ represents
comoving distances, $x$ represents physical distances, and $\Phi$ is
the gravitational potential. For simplicity, we are restricting
ourselves throughout this section to a flat Universe and a flat sky
approximation; for a curved sky the calculation can be extended using
the formalism of Castro et al (2005). The equation above for
convergence can be put into terms of the overdensity of matter using
the Poisson equation

\begin{equation}
\frac{\partial^2}{\partial x_i \partial x_i}
\Phi \simeq \frac{3 H_0^2 \Omega_m \delta}{2a}
\end{equation}
which gives:

\begin{equation}
\kappa({\bf \theta},w)=\frac{3 H_0^2 \Omega_m}{2c^2}\int_0^w dw'
\frac{(w-w')w'}{w}\frac{\delta[w'{\bf \theta}, w']}{a(w')}.
\label{eqn:kappa}
\end{equation}
Now we wish to differentiate this to obtain a form for the effective
cosmological flexion. In order to do this, we note the relationship
between the required gradient with respect to angle on the sky, and
the gradient of physical distances:

\begin{equation}
\partial_i=w\frac{\partial}{\partial x_i}.
\end{equation}
Using this, we obtain for the first flexion

\begin{eqnarray}
\flex=\partial_i \kappa&=& \frac{3 H_0^2 \Omega_m}{2c^2}\int_0^w dw'
\frac{(w-w')w'^2}{a(w') w}\frac{\partial}{\partial x_i} \delta[w'{\bf
\theta}, w']\nn
&=&\frac{3 H_0^2 \Omega_m}{2c^2}\int_0^{w_H} dw
\frac{\bar{W} w^2}{a(w)} \delta'[w{\bf
\theta}, w].
\label{eqn:feff2}
\end{eqnarray}
Here, $\delta'$ is the transverse gradient of the overdensity, and we have defined

\begin{equation}
\bar{W}=\int_w^{w_H} dw' G(w') \frac{(w'-w)}{w'}
\end{equation}
where $G$ is the distribution of galaxies as a function of radial distance.

\begin{figure}
\psfig{figure=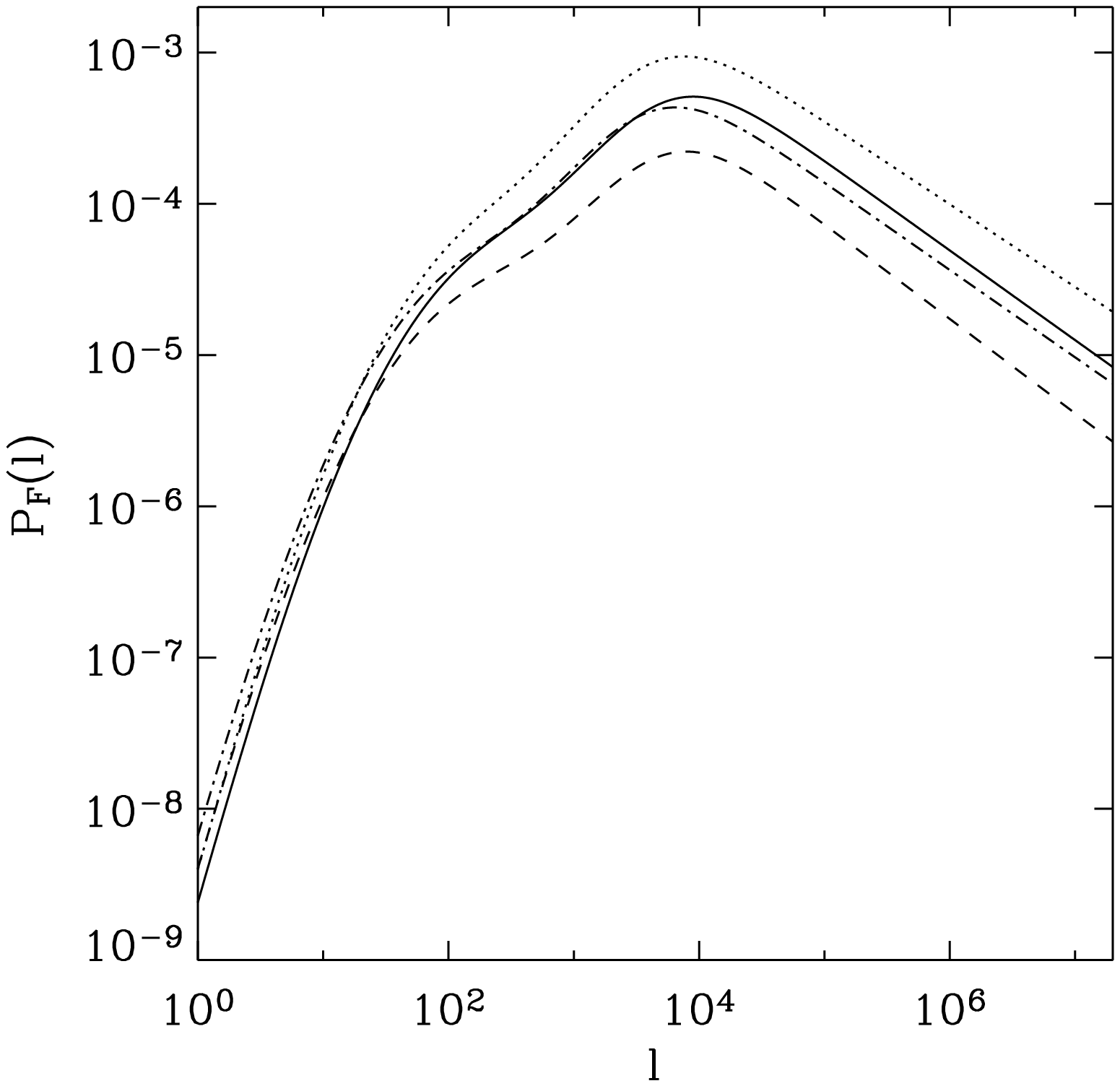,height=7cm,angle=0}
\psfig{figure=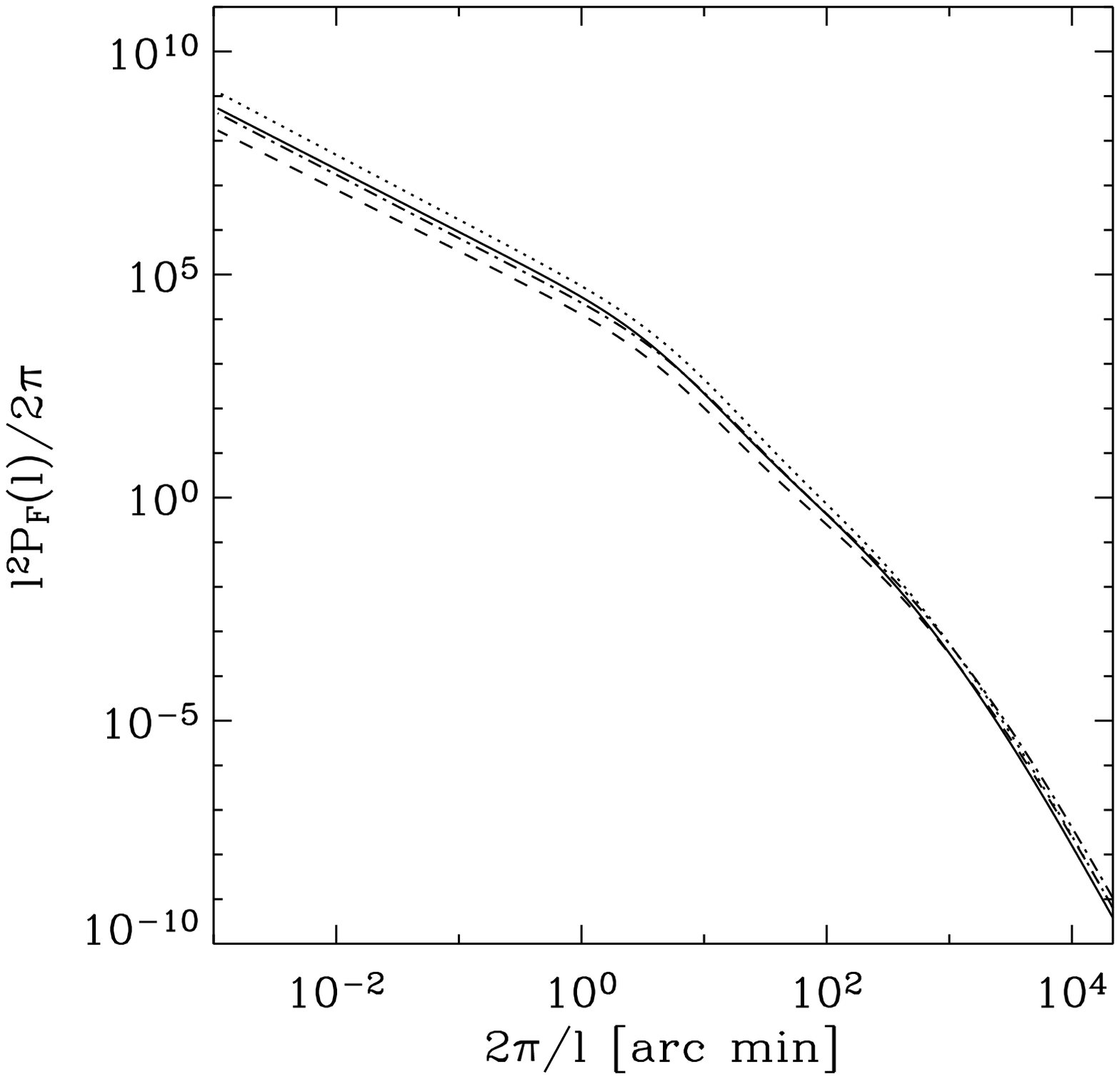,height=7cm,angle=0}
\caption{The cosmic flexion power spectrum. Top: power spectrum as a
function of angular wavenumber $l$; bottom: power spectrum per log
interval in angular scale. Solid line: $(\Omega_m=0.3,
\Omega_\Lambda=0.7,\sigma_8=0.7)$; dotted line: $(\Omega_m=0.3,
\Omega_\Lambda=0.7,\sigma_8=0.9)$; dashed line: $(\Omega_m=0.2,
\Omega_\Lambda=0.8,\sigma_8=0.7)$; dash-dotted line: $(\Omega_m=0.2,
\Omega_\Lambda=0.8,\sigma_8=0.9)$.}
\end{figure}

\begin{figure}
\psfig{figure=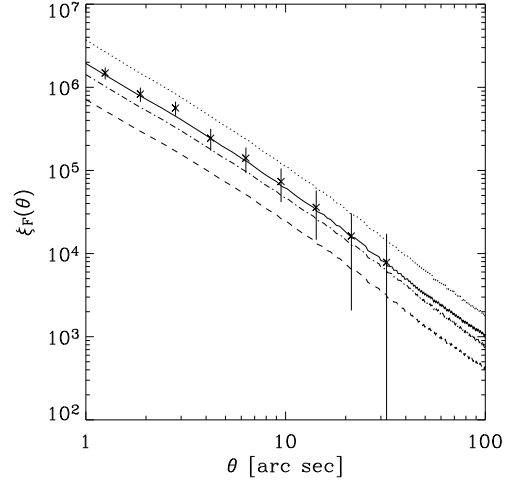,height=7cm,angle=0}
\caption{Cosmic flexion correlation function for the cosmological
models shown in Figure 9. Also plotted is the error on cosmic flexion
for a 100 square degree ground-based survey.}
\end{figure}

In order to find the power spectrum of cosmic flexion, we will use a
form of Limber's equation, which states that if one can find two
quantities $g_1$ and $g_2$ written in terms of some other
quantities $q_i$ as

\begin{equation}
g_i=\int dw' q_i(w') G(w') \delta'[w'{\bf
\theta}, w']
\end{equation}
then the cross-power spectrum of $g_1$ and $g_2$ is

\begin{equation}
P_{12}(\ell)=\int dw' \frac{q_1(w') q_2(w')}{w'^2} P_{\delta'}(k,w'),
\label{eq:limber}
\end{equation}
where $\ell$ is the angular wavenumber and $P_{\delta'}$ is the power
spectrum of the transverse gradient of the density fluctuations.  We
note that we can write the flexion in equation (\ref{eqn:feff2}) in
this way, with $q$ given by

\begin{equation}
q=\frac{3H_0^2 \Omega_m}{2 c^2} \frac{\bar{W}(w) w^2}{a(w)}.
\end{equation}
Therefore we can write the flexion power spectrum as

\begin{equation}
P_{\flex}(\ell) = \frac{9H_0^4 \Omega_m^2}{4 c^4} \int dw \frac{\bar{W}^2(w)
w^2}{a^2(w)} P_{\delta'} \left( \frac{\ell}{w},w \right).
\end{equation}
Because flexion is the derivative of convergence, this power spectrum
is in terms of the derivative of the overdensity. In order to describe
the flexion power spectrum in terms of the more familiar overdensity
itself, we note that

\begin{equation}
|\delta_k'|^2 = |\delta_k|^2 k_1^2.
\end{equation}
This implies that

\begin{equation}
P_{\delta'} \left( \frac{\ell}{w},w \right)=P_{\delta} \left(
\frac{\ell}{w},w \right) \frac{\ell^2}{w^2}.
\end{equation}
Finally, then, we can describe the flexion power spectrum as

\begin{equation}
P_{\flex}(\ell) = \frac{9H_0^4 \Omega_m^2}{4 c^4} \ell^2 \int dw
\frac{\bar{W}^2(w)}{a^2(w)} P_{\delta} \left( \frac{\ell}{w},w \right).
\label{eqn:pf}
\end{equation}
We note that this has a very similar form to the convergence power
spectrum, differing only by a factor of $\ell^2$. Thus flexion power will
be dominated by high $\ell$ components; again we see that flexion takes
the form of a high-bandpass filter for density fluctuations.

One can easily show that the two-point statistics of $\flex$ and
$\sflex$ are identical; hence the first flexion power spectrum which
we have calculated here is identical to the second flexion power
spectrum.

From this power spectrum, we can find the flexion correlation
function, as these are related by:

\begin{eqnarray}
\xi_{\flex}(\theta) = \int_0^\infty \frac{d^2 \ell}{(2 \pi)^2}
P_{\flex}(\ell) e^{-i{\bf \ell.\theta}}\nn =\int_0^\infty \frac{\ell
d\ell}{2 \pi} P_{\flex}(\ell) J_0(\ell\theta).
\end{eqnarray}
Thus

\begin{eqnarray}
\xi_{\flex}(\theta) = \frac{9H_0^4 \Omega_m^2}{4 c^4} \int_0^{w_H} dw
\frac{\bar{W}^2(w) w^4}{a^2(w)} \nn \times \int_0^\infty \frac{k
dk}{2\pi}P_{\delta}(k,w) k^2 J_0(k w \theta).
\end{eqnarray}
We can now examine what these predictions provide in practice. We
numerically calculate the flexion power spectrum from equation
(\ref{eqn:pf}) using the matter power spectrum prescription used in
Bacon et al. (2004). This uses an initial Harrison-Zel'dovich power
spectrum with non-linear evolution following Smith et al. (2003).

Figure 9 shows the flexion power spectrum in two forms. In the top
panel, we present the power spectrum as a function of angular
wavenumber $l$, for median redshift $z=1$. It is clear that the
flexion power predictions are significantly dependent on the
cosmological model; we will discuss whether this affords measurement
of cosmological parameters in the context of correlation functions
below.  We note that the flexion power peaks at smaller angular scales
than the shear power spectrum, i.e. $\sim 1$ arcmin as opposed to a
few 100 arc min (c.f. Bartelmann \& Schneider 2001, Figure 16). We
also note that the flexion power spectrum has a very familiar shape;
since the shear power spectrum is often shown premultiplied by
$\ell^2$, the flexion power spectrum (without premultiplication by
$\ell^2$) is identical in shape to the premultiplied shear power
spectrum.

The bottom panel shows the flexion power per logarithmic interval in
angular wavenumber. This shows that, for reasonable cosmological
models, the power per log interval increases without limit for Smith
et al (2003) density spectra. This is in contrast to the shear power
spectrum, where one finds a broad maximum in power per log interval
below $\simeq 1'$ (c.f. Bartelmann \& Schneider Figure 16). This again
illustrates that cosmic flexion is increasingly sensitive to dark
matter concentrations on small scales.

Figure 10 shows predictions for the cosmic flexion correlation
function for median redshift $z=1$, where we plot flexion in units
rad$^{-1}$. Note again the significantly different predictions for
different cosmologies. However, we also plot errors in measuring the
cosmic flexion, for a 100 square degree ground-based survey with
galaxy number density of 20 arcmin$^{-2}$. Note that these error bars
will have significant covariance between angular scales. We see that,
while on small scales we can obtain a clear measurement of the
small-scale structure, we cannot obtain measurements of the flexion in
the linear density regime. This makes cosmological parameter
prediction unfeasible, as it is difficult to predict amplitudes for
structure on very nonlinear scales from cosmological
models. Nevertheless, cosmic flexion is useful in probing these scales
in order to understand them on their own terms, describing
substructure and the cuspiness of halos; cosmic flexion is also
complementary to cosmic shear, probing small scales in an isolated
fashion, whereas cosmic shear has a broad window function for
power. The cosmic flexion signal will be a useful means of testing
theories of stable clustering or stable merging (c.f. Smith et al
2003).

It should be noted that in this analysis we have neglected the power
that might exist from intrinsic, physical flexion correlations between
galaxies. The analogous intrinsic ellipticity correlation between
galaxies has been shown (e.g. Heymans et al 2004) to be small;
however, further work will be necessary to measure the level of
contamination of cosmic flexion due to intrinsic flexion alignments.

\subsection{Convergence-Flexion Cross Power Spectrum}

\begin{figure}
\psfig{figure=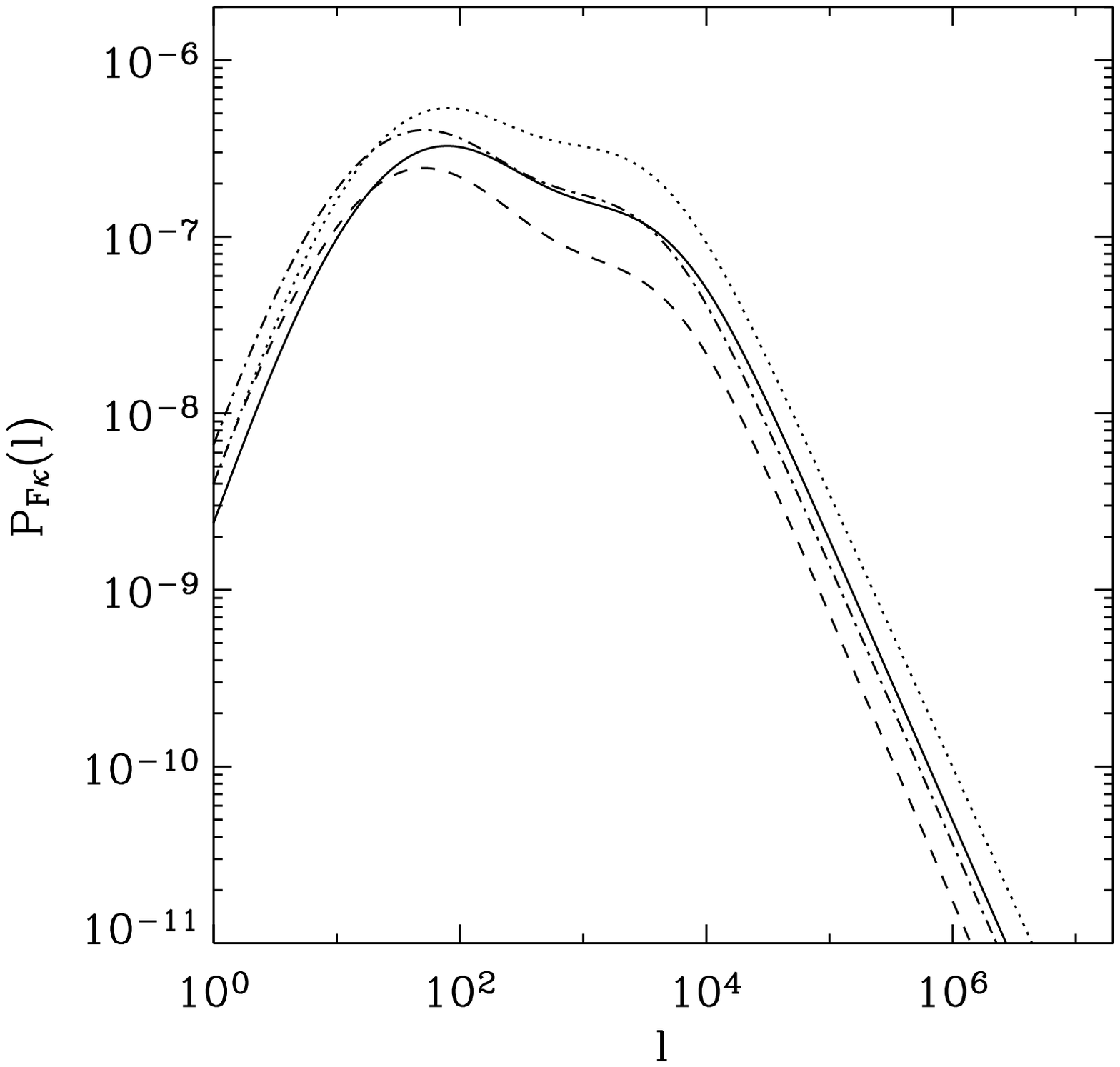,height=7cm,angle=0}
\psfig{figure=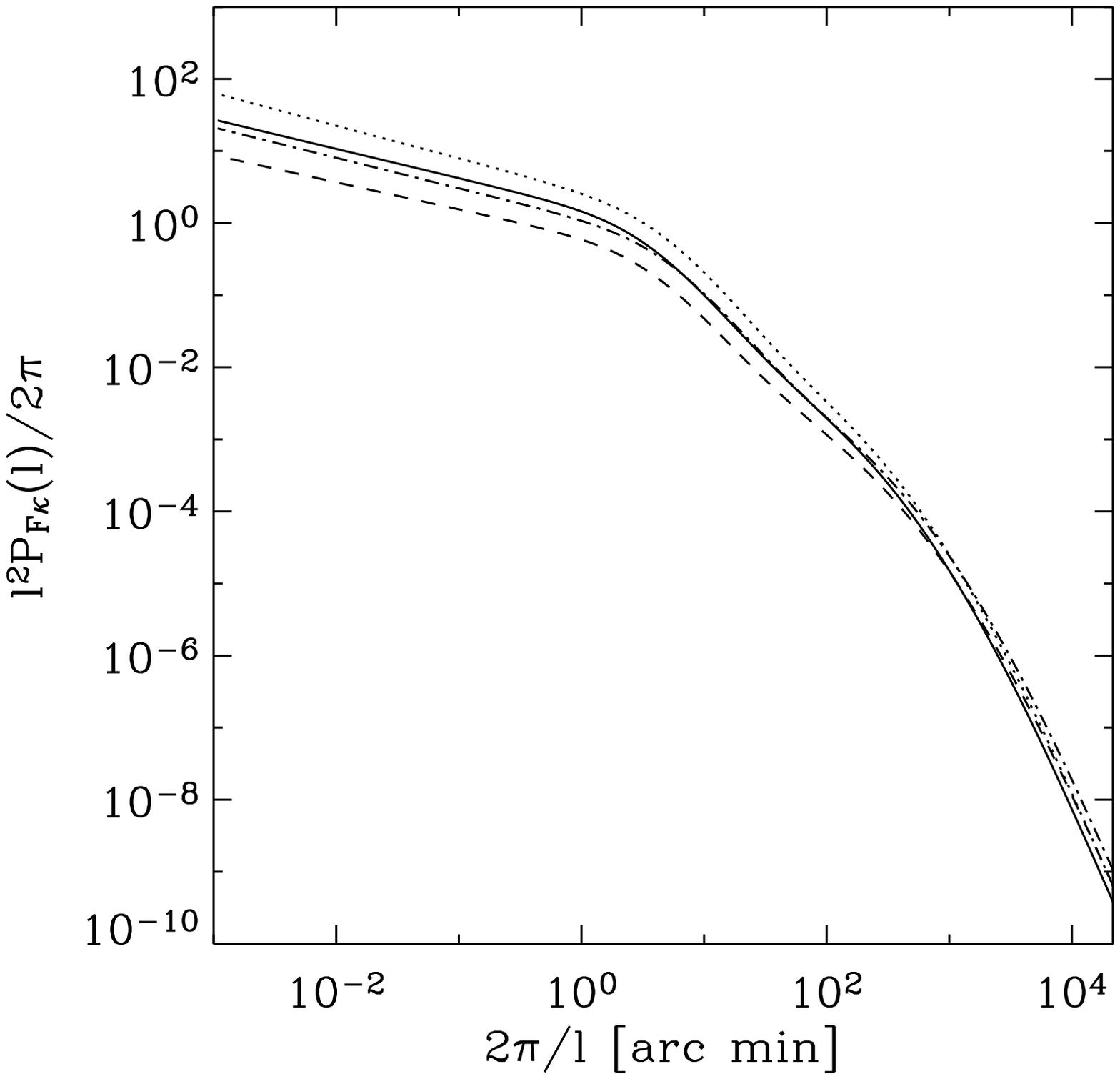,height=7cm,angle=0}
\caption{The cosmic convergence-flexion cross-power spectrum. Top: power
spectrum as a function of angular wavenumber $l$; bottom: power
spectrum per log interval in angular scale. The lines represent the
same cosmological models as in Figure 9.}
\end{figure}

\begin{figure}
\psfig{figure=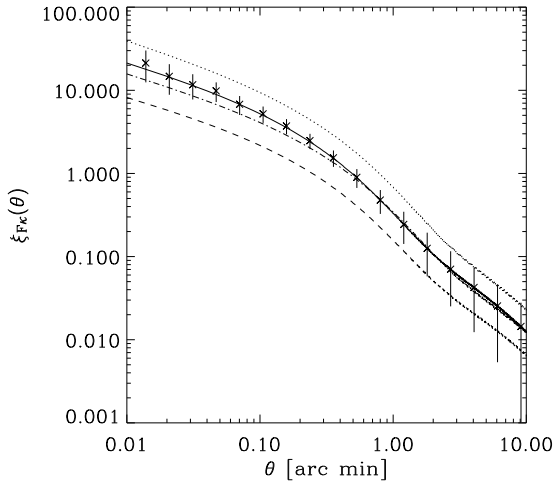,height=7cm,angle=0}
\caption{Cosmic convergence-flexion cross-correlation function for the
cosmological models shown in Figure 9. Also plotted is the error on
cosmic flexion-convergence cross-correlation for a 100 square degree
ground-based survey.}
\end{figure}

In addition to the flexion power spectrum, we are also able to
calculate the convergence-flexion cross power spectrum, which can
easily be related to the shear-flexion cross power spectrum. We note
that to do this we can again use Limber's equation (\ref{eq:limber}),
but this time using $P_\delta$ from the outset rather than
$P_{\delta'}$. In this case, from our final power spectrum for flexion
(equation \ref{eqn:pf}) we see that the relevant choice of $q$ for
flexion in Limber's equation is

\begin{equation}
q_{\cal F} = \frac{3H_0^2 \Omega_m \bar{W}(w) w \ell}{2c^2a(w)}.
\end{equation}
In addition, from equation (\ref{eqn:kappa}), we see that the choice
of $q$ suitable for convergence is

\begin{equation}
q_\kappa = \frac{3H_0^2 \Omega_m \bar{W}(w) w}{2c^2 a(w)}.
\end{equation}
Hence the cross-power spectrum between convergence and flexion can be
written as

\begin{equation}
P_{{\cal F}\kappa}(\ell)=\frac{9H_0^4\Omega_m^2}{4c^4} \int dw
\frac{\bar{W}^2(w)}{a^2(w)} P_\delta \left(\frac{\ell}{w},w \right) \ell.
\end{equation}
This is shown in Figure 11, together with the associated
convergence-flexion cross-correlation function in Figure 12 with
appropriate errors for a 100 square degree survey. We see that this
quantity has a measurement limit on an intermediate scale to shear and
flexion limits ($\simeq 2'$). It is a valuable quantity to measure, as
it gives a stronger signal-to-noise than cosmic flexion, and offers a
stringent check on systematic errors between the shear or convergence
and flexion signals.

\section{Conclusions}

In this paper, we have examined how flexion can be applied to obtain
both astrophysical and cosmological information. We have explored the
use of galaxy-galaxy flexion to measure the mass and profile of galaxy
dark matter halos; we have shown how flexion can generate maps of dark
matter; and we have calculated the cosmic flexion correlation signal.

We have presented a flexion formalism, showing how the effect arises
from the variation of the shear field over an object, and giving a
brief discussion of how the effect can be measured using shapelets.  A
second flexion which was not considered in previous work has also been
presented; this second flexion contains non-local information which
generates arcs from point mass lenses, while the first flexion
contains local information about the gradient of the density.

We have examined the efficiency of flexion as a description of
second-order lensing information, in comparison with simply describing
this in terms of gradients of shear. Flexion is found to be an
optimal description for point mass lensing, and is about as efficient
as shear gradients for singular isothermal spheres.

We have calculated flexion predictions for galaxy-galaxy lensing, for
a variety of galaxy halo profiles including the singular isothermal
sphere, with or without softening, the elliptical isothermal, and the
NFW profile. It is found that galaxy mass can be measured well with
flexion, as the mass-sheet degeneracy which plagues shear does not
exist for flexion. Also, we find that by combining shear and flexion
galaxy-galaxy lensing, we are able to produce powerful constraints on
the halo profile.

Flexion can be used to reconstruct mass profiles directly, using a
similar process to the Kaiser-Squires (1993) and Taylor (2001)
inversions in 2 and 3 dimensions respectively. We have noted how
flexion can act as an excellent tool for measuring substructure.

We have also calculated predictions for cosmic flexion, the flexion
arising from large-scale structure. It is found that this signal is only
measurable on small scales; it is useful for measuring small-scale
structure and halo profiles, but will not yield independent
cosmological parameters, as predictions for structure
amplitudes are difficult in this highly non-linear regime.

We have seen from these applications of flexion that this quantity is
a highly useful tool for a variety of methods of measuring mass fluctuations
in the Universe. Flexion constitutes a valuable complement to shear,
as it is sensitive where shear is not, and vice versa. With upcoming
surveys from ground and space, flexion will provide a useful addition
to the armoury of those who seek to understand mass in the Universe.

\section*{Acknowledgments}

DJB and ANT are supported by PPARC Advanced Fellowships. DMG is
supported by NASA ATP Grant \#NNG05GF616.  We would like to thank John
Peacock, Peter Schneider and Tereasa Brainerd for very useful
discussions.

\begin{appendix}
\section{NFW halo parameter conventions}
We will follow the lead of
Kleinheinrich et al. (2005) and briefly discuss the differing
conventions used to describe NFW halos in the literature. In this
comparison, and in the section above, we have adopted the convention
used by Navarro et al. (1996, 1997) and by Hoekstra et al. (2004)
of defining a radius $r_{200}$ from the centre of a CDM
halo within which the mean density is 200 times the
\emph{critical} density for closure of the universe in that epoch.
The mass of the halo can then be quantified via $M_{200}$, the mass
contained within $r_{200}$ such that
\begin{equation}
M_{200} = \frac{800 \pi}{3}\rho_{crit}(z) r^3_{200}.
\end{equation}
The scaling radius $r_s$ of equation (\ref{nfw}) is then expressed by
Navarro et al. (1997) in terms of $r_{200}$ and another dimensionless
scaling parameter, the concentration $c$, as $r_s = r_{200}/c$.
From the definition of $M_{200}$, the parameters $c$ and $\Delta_c$ are
linked by the relation
\begin{equation}
\Delta_c = \frac{200}{3} \frac{c^3}{[\ln(1+c) - c/(1+c)]}.
\end{equation}
The convention outlined above is not used by all authors, with
Kleinheinrich et al. (2005) choosing to define $r_{200}$ as the radius
from the halo centre within which the mean density is 200 times the
overall mean \emph{matter} density of the universe at that epoch.
This convention, which we will hereafter denote via the use of primes,
thus relates $M_{200}'$ to $r_{200}'$ through
\begin{equation}
M_{200}' = \frac{800 \pi}{3} \Omega_m(z)\rho_{crit}(z)r_{200}'^3
\end{equation}
where $\Omega_m(z)$ is the matter density parameter at
the epoch of the halo in question. For any given halo at a redshift
$z$ we can hence define a concentration $c'$ such that $r_s =
r_{200}'/c'$ and a characteristic density related to the concentration
as follows:
\begin{equation}
\Delta_c' = \frac{200\,\Omega_m(z)}{3}\frac{c'^3}{[\ln(1+c')-c'/(1+c')]}.
\end{equation}
We note that while $M_{200}'$, $r_{200}'$ and $c'$ take
different values to their unprimed counterparts, $r_s$ must not change
and we must have $\Delta_c = \Delta_c'$, as both these parameters describe
the real physical density profile of the halo.

Given the potential for confusion of having two differing NFW
conventions in the literature, it is worthwhile to describe the
conversion between the two. If we have a halo of concentration $c$,
defined as by Navarro et al. (1997), at a redshift $z$, then it can be
quickly seen that the corresponding concentration for the primed
convention is found by solving
\begin{equation}
\frac{\Omega_m(z) \, c'^3}{[\ln(1+c')-c'/(1+c')]} =
\frac{c^3}{[\ln(1+c)-c/(1+c)]}.
\end{equation}
Once $c'$ is determined, the conversion relations for $r_{200}'$
and $M_{200}'$ follow trivially:
\begin{equation}
\frac{r_{200}'}{r_{200}}=\frac{c'}{c}, \;\;\;\;\;\;\;\;
\frac{M_{200}'}{M_{200}}=\Omega_m(z)\left(\frac{c'}{c}\right)^3.
\end{equation}
Finally we note that in practice the primed values of $c'$, $M_{200}'$
and $r_{200}'$ are somewhat larger than their unprimed counterparts.
\end{appendix}

\label{lastpage}

\end{document}